\documentclass[AMA,STIX1COL]{WileyNJD-v2}
\articletype{Article Type}%
\usepackage{amsmath}
\usepackage{graphicx}
\usepackage{textcomp}
\usepackage{float}
\usepackage{listings}
\usepackage{booktabs}
\usepackage{subfig}
\usepackage{mdwlist}
\usepackage{array}
\usepackage{subfiles}
\usepackage{import}
\usepackage{setspace}
\usepackage{dsfont}
\usepackage{mathtools}
\usepackage{xcolor}
\usepackage{sjmacros}

\begin{document}
\theoremstyle{definition}
\newtheorem{example}{Example}[]

\newcommand{\itgrlexp}[1]{\frac{1}{#1^2}\left(1 - e^{-#1t}\left(1 + #1t\right)\right)}
\newcommand{\itgrlexpparens}[1]{\frac{1}{(#1)^2}\left(1 - e^{-(#1)t}\left(1 + (#1)t\right)\right)}

\thispagestyle{empty}
\allowdisplaybreaks
\raggedbottom

\title{Incremental Nonlinear Stability Analysis of Stochastic Systems Perturbed by L\'{e}vy Noise}

\author[1]{SooJean Han*}
\author[1,2]{Soon-Jo Chung}
\authormark{Han \textsc{et al}}

\address[1]{\orgdiv{Department of Computing and Mathematical Sciences}, \orgname{California Institute of Technology}, \orgaddress{Pasadena, \state{California}, \country{USA}}}

\address[2]{\orgdiv{Graduate Aerospace Laboratory}, \orgname{California Institute of Technology}, \orgaddress{Pasadena, \state{California}, \country{USA}}}

\corres{*Corresponding author, \email{soojean@caltech.edu}}


\abstract[Summary]{
We present a theoretical framework for characterizing incremental stability of nonlinear stochastic systems perturbed by compound Poisson shot noise and finite-measure L\'{e}vy noise. 
For each noise type, we compare trajectories of the perturbed system with distinct noise sample paths against trajectories of the nominal, unperturbed system. We show that for a finite number of jumps arising from the noise process, the mean-squared error between the trajectories exponentially converge towards a bounded error ball across a finite interval of time under practical boundedness assumptions.
The convergence rate for shot noise systems is the same as the exponentially-stable nominal system, but with a tradeoff between the parameters of the shot noise process and the size of the error ball.
The convergence rate and the error ball for the L\'{e}vy noise system are shown to be nearly direct sums of the respective quantities for the shot and white noise systems separately, a result which is analogous to the L\'{e}vy-Khintchine theorem.
We demonstrate our results using several numerical case studies.
}


\maketitle


\section{Introduction}
\subsection{Motivation and Related Work}
Many model-based controllers or observers for robotics applications are typically designed for robustness against additive Gaussian white noise. 
Gaussian white noise systems are appealing to study because they demonstrate properties (e.g., independent increments, Central Limit Theorem) which make them easier and more convenient to analyze than non-Gaussian stochastic systems.
The choice to study Gaussian white noise processes is also justifiable in practice; vision-based localization and mapping~\cite{yang13}, spacecraft navigation~\cite{capuano19}, and motion-planning~\cite{kalakrishnan11} are notable examples of robotics applications which employ the Gaussian white noise model.
Consequently, there has been a wealth of literature devoted towards the study of Gaussian white noise systems, particularly in stability analysis, controller, and observer design.
For example, a model-based controller can be developed via the well-known Linear Quadratic Gaussian (LQG) approach~\cite{doyle78}, and a model-based observer can be designed via Kalman filtering and its extensions~\cite{reif99,wan00}.
More recent methods of model-based controller synthesis for Gaussian white noise systems include the path integral approach~\cite{theodorou10}, convex optimization-based approaches~\cite{rawlik13,nakka19}, as well as a number of reinforcement learning-based approaches~\cite{munos98,deisenroth15}.

However, there is a major lack of generality with Gaussian white noise, which makes them unsuitable for use in non-Gaussian problems, e.g., fault diagnosis and fault-tolerant control~\cite{yao12}, filtering when observations are received unreliably~\cite{battilotti19six}, and target-tracking with nonlinear measurement equation~\cite{cacace19three}.
In many cases, machine-learning-based methods~\cite{shi20,tsukamoto20neural,nakka21} are the go-to approaches used to perform control or estimation in non-Gaussian noise stochastic systems.
But a downside of model-free methods are the large amounts of time and training data needed to learn information that could be obtained using structured models.
On the other hand, training time and data can be reduced by expanding the set of model assumptions and considering a broader class of noise distributions prior to applying model-free techniques.
For example, Cacace 2019~\cite{cacace19two} explores the possibility of developing more accurate estimators for linear non-Gaussian systems beyond the Kalman filter, which is only optimal for linear Gaussian systems and when estimators are constrained to be affine.

However, because the class of non-Gaussian noise processes is diverse, verifying stability of non-Gaussian stochastic systems is an important challenge to address prior to considering observer or controller design; Battilotti 2019~\cite{battilotti19one} highlights this even for the case of linear dynamics.
For systems modeled as right-continuous, strong Markov processes, the seminal work of Kushner 1967~\cite{kushner67} laid out the foundations of stochastic stability theory by considering the following question: can trajectories of the system, arising from different sample paths of noise and different initial conditions, be bounded in some region after a sufficiently elapsed time?
The common Lyapunov approach addresses this question for when the bounded region is an equilibrium point or a limit cycle.
Alternative approaches to characterize stability for non-Gaussian stochastic systems have also been studied in literature: asymptotic stability of systems driven by L\'{e}vy noise is developed in Applebaum 2009~\cite{applebaum09} while exponential stability for systems perturbed by semimartingale processes is studied in Mao 1990~\cite{mao90}.

Incremental stability~\cite{lohmiller00,angeli02} considers the stronger condition in which multiple distinct solution trajectories converge globally exponentially towards each other.
Applications of incremental stability arise in numerous settings such as cooperative control over multi-agent swarm systems~\cite{chung09} and phase synchronization in directed networks~\cite{slotine04,chung13}. 
A recent tutorial paper on incremental stability and connections to machine learning is Tsukamoto et. al. 2021~\cite{tsukamoto21}.
For trajectories of incrementally stable stochastic systems with distinct initial conditions and distinct sample paths of noise, global exponential convergence towards a reference or nominal trajectory is guaranteed to within a bounded error ball.
Pham 2009~\cite{pham09} derived incremental stability conditions for stochastic systems perturbed by additive Gaussian white noise.
Dani 2015~\cite{dani15} extends Pham 2009 to more general state-dependent contraction metrics, then considers the problem of observer design for Gaussian white noise systems. 
However, to the authors' knowledge, incremental stability has not been studied for any distribution of non-Gaussian stochastic system.

One especially prevalent distribution of non-Gaussian noise is impulsive \textit{shot noise}~\cite{baccelli_book}, which arises in real-world applications almost just as frequently as Gaussian white noise does.
Some examples are the signal-processing neuronal spikes arising from brain activity in the field of neuroscience~\cite{patel09}, large fluctuations in stock prices in the field of economics~\cite{cont06}, and in impulsive noisy sensor measurements in spacecraft control~\cite{raghunathan98}.
In the field of robotics, impulsive noise can arise as large proprioceptive measurement errors~\cite{pan15} or disturbances due to obstacle collisions or wind turbulence~\cite{oconnell20}.
Despite this prevalence, there are little to no model-based synthesis procedures for controllers or observers dedicated to robustness against impulsive shot noise. 
Because Gaussian white noise is usually small in magnitude and continuous in the sense that changes occur over a measurable duration of time, it cannot be used to model sudden impulsive perturbations like shot noise.

In the field of applied mathematics, there is an abundance of literature which provides useful tools and theoretical results that can be used to model the shot noise phenomenon. 
Just as how the standard Brownian motion process is used to model various forms of Gaussian white noise, shot noise is modeled using the \textit{Poisson processes}, especially the \textit{compound Poisson process}.
Furthermore, like Brownian motion processes, Poisson processes have stationary and independent increments.
In fact, both Brownian motion and Poisson processes are special cases of the more general \textit{L\'{e}vy processes} which have these shared properties. 
A particularly useful result is the \textit{L\'{e}vy-Khintchine Theorem}, which describes L\'{e}vy processes as affine combinations of Brownian motion processes and compound Poisson processes~\cite{applebaum09_book}. 

\subsection{Our Contributions}
%
This paper uses the theory of Poisson random measures and L\'{e}vy processes to develop incremental stability criteria for stochastic nonlinear systems perturbed by two types of non-Gaussian noise.
The first type of noise process are shot noise processes which are represented as compound Poisson processes. 
The second are finite-measure L\'{e}vy processes constructed by taking the affine combination of Gaussian white and compound Poisson shot noise processes. 
To address stochastic incremental stability for each type of system, we use contraction theory to compare trajectories of the noise-perturbed system with distinct noise sample paths against trajectories of the nominal, unperturbed system, starting from different initial conditions. 
We show that when a finite number of jumps arise from the noise process over a fixed, finite interval of time, the mean-squared error between the trajectories converge exponentially to a bounded error ball under the assumption that certain parameters of the noise process and contraction metric are bounded.
%
The convergence rate and the error ball for the L\'{e}vy noise system is shown to be nearly a direct summation of the respective quantities for the shot and white noise systems separately, in likeness to the statement of the L\'{e}vy-Khintchine Theorem. 
For concreteness, we specialize our results to linear time-varying (LTV) dynamics, which allows us to use an explicit form of contraction metric and provide additional insights to our general results.
We qualitatively discuss how both 1) fixed, inherent parameters of the system (e.g., intensity and maximum norm of jumps in the noise process) and 2) design parameters (e.g., contraction metric) affect the tightness of the stability bounds.
To illustrate our results, several empirical studies are performed for various systems:
a 1D linear reference-tracking system, 2D linear time-varying systems, and a 2D nonlinear system.

\subsection{Paper Organization}
We begin in~\sec{preliminaries} by outlining notation and briefly reviewing background.
In~\sec{problem_setup}, we establish the form of shot and L\'{e}vy noise system dynamics used throughout the paper, and set up the incremental stability background needed for our main results.
Sections~\ref{sec:stoch_contract} and~\ref{sec:simulation} present our main results.
In Section~\ref{subsec:stoch_contract_thms}, we state and prove our two theorems, the Shot Contraction Theorem and the L\'{e}vy Contraction Theorem. 
Section~\ref{subsec:shot_contract_linear} specializes the Shot Contraction Theorem to linear time-varying (LTV) nominal systems.
In~\sec{simulation}, we use numerical case studies to supplement the theory developed in~\sec{stoch_contract}.
Section~\ref{subsec:2D_nonlinear} demonstrates the Shot Contraction Theorem for a  simple 2D nonautonomous, nonlinear system.
Section~\ref{subsec:1D_lqr} considers reference-tracking for a simple 1D linear shot noise system, and Section~\ref{subsec:2D_ltv} extends this to two different 2D LTV systems. 
Finally, we conclude the paper in~\sec{conclusion}.


\section{Preliminaries}\label{sec:preliminaries}
\subsection{Terminology and Mathematical Notation}\label{subsec:notation}
For the sake of simplifying terminology, we refer to additive Gaussian white, additive compound Poisson shot, and additive bounded-measure L\'{e}vy noise systems as simply white, shot, and L\'{e}vy noise systems, respectively.
The understanding is that the shortened terminology throughout this paper does not refer to the general cases (e.g., non-Gaussian white noise, or L\'{e}vy noise whose measures have infinite total mass).
We use the $\ell_p$ norm $\norm{\cdot} \triangleq \norm{\cdot}_p$ for vectors and the corresponding induced norm $\norm{A} \triangleq \sup_{\xvect\neq 0} (\norm{A\xvect}/\norm{\xvect})$ for matrices. We abuse the notation $\norm{\cdot}$ to apply to both matrices and vectors where the context is clear.
For any $g$ which is a function of time, we denote the left-limit of $g(t)$ at time $t$ as $g(t-) \triangleq \lim_{s\to t}g(s)$ where $s<t$ for any $t > 0$.
For any $g:\Rbb\times\Rbb^n\to\Rbb^m$, a function of both a scalar parameter $a\in\Rbb$ and a vector $\xvect\in\Rbb^n$, we write the partial derivatives in the following shorthand notation. First, $\partial_ag \equiv \partial_ag(a,\xvect) \triangleq \partial g(a,\xvect)/\partial a$, and $\nabla_{\xvect}g \equiv \nabla_{\xvect}g(a,\xvect) \in \Rbb^{m\times n}$ denotes the gradient with respect to $\xvect$. Moreover, we denote $\partial_{x_ix_j}^2 \triangleq \partial^2 g(a,\xvect)/(\partial x_i\partial x_j)$ as the double partial derivative of $g$ with respect to two distinct components $x_i$ and $x_j$ of $\xvect$, $i\neq j$, and likewise $\partial_{x_i}^2 \triangleq \partial^2 g(a,\xvect)/(\partial^2 x_i)$ for the derivative with respect to the same component $x_i$.
When describing a dynamical system of the form $\dot{\xvect}(t) = f(t,\xvect(t))$, where $\xvect:\Rbb^{+}\to\Rbb^n$ and $f:\Rbb^{+}\times\Rbb^n\to\Rbb^n$, we simplify the notation by writing $\xvect(t)$ without the argument $t$, as in $\dot{\xvect} = f(t,\xvect)$ with the understanding that $f$ is not taking the function $\xvect$ as input, but rather the vector $\xvect(t)$.
We define the dot notation as the time-derivative, meaning $\dot{\xvect}(t) \triangleq (d/dt)\xvect(t)$.
Any function $H(t,\xvect(t))$ with arguments time $t\in\Rbb^{+}$ and state $\xvect(t)\in\Rbb^n$, which evolves over time according to some dynamics of the form $\dot{\xvect} = f(t,\xvect)$, has the time-derivative $\dot{H} \triangleq \partial_tH + \nabla_{\xvect}H\cdot\dot{\xvect}(t) = \partial_tH + \nabla_{\xvect}H\cdot f(t,\xvect)$.

\subsection{Poisson and L\'{e}vy Processes}\label{subsec:bg_assump}
The general \textit{Poisson random measure} is characterized by an \textit{intensity measure} Leb$\times\nu$, where Leb denotes the standard Lebesgue measure over time and $\nu(dy)$ is the probability measure over a space $E$ describing the distribution of jumps. 
Such a Poisson random measure is typically denoted $N(dt,dy)$ over the space $[0,t]\times E$.
While most results in the theory of Poisson processes are written with respect to general Poisson random measures, our scope in this paper is specifically on the standard and compound Poisson processes, defined below.

\begin{definition}[Standard and Compound Poisson Processes]\label{def:cmpd_poisson}
    Let $E \subseteq \Nbb^{\ell}$ and $t > 0$. 
    The \textit{standard Poisson process} $N(t)$ counts the number of jumps in $E$ that have occurred in the time interval $[0,t]$ for $t\leq T$. 
    It is characterized by an \textit{intensity parameter} $\lambda > 0$, which describes the average rate at which jumps occur in the process.
    The \textit{compound Poisson process} $\sum_{i=1}^{N(t)} \xi(T_i)$ is a simple generalization of the standard Poisson process to include weighted jumps, where $T_i < t$ is the arrival time of the $i$th jump and $\xi:[0,T]\to E$ is a function describing the jump distribution over the space $E$. 
    The intensity $\lambda > 0$ of a compound Poisson process is the same as its corresponding standard Poisson process $N(t)$.
\end{definition}

\begin{definition}[L\'{e}vy Processes]\label{def:levy_process}
    A process $\left\{ L(t) : t\geq 0\right\}$ defined on a probability space $(\Omega,\Fcal,\Pbb)$ is said to be a \textit{L\'{e}vy process} if the following hold:
    \begin{itemize}
        \item C\'{a}dl\'{a}g Paths: the paths of $L$ are almost-surely $\Pbb$ right-continuous with left-limits. This means that every sample path of $L$ must be right-continuous with left-limits.
        \item Zero Initial Condition: $\Pbb(L(0) = 0) = 1$
        \item Stationary Increments: for $0\leq s\leq t$, $L(t) - L(s)$ is equal in distribution to $L(t-s)$.
        \item Independent Increments: for $0\leq s\leq t$, $L(t) - L(s)$ is independent of $L(r)$ for $r \leq s$.
    \end{itemize}
\end{definition}

\begin{remark}\label{rmk:levy_khintchine}
    Under~\defin{levy_process}, we can observe that both Gaussian white noise processes and compound Poisson shot noise processes are L\'{e}vy processes.
    This implies that the affine combination of the two is also a L\'{e}vy process. 
    In fact, a well-known result called the \textit{L\'{e}vy-Khintchine Theorem} (see, e.g., Theorem 1.2.14 of Applebaum 2009~\cite{applebaum09}) says that L\'{e}vy processes can be represented as affine combinations of Brownian motion processes and Poisson processes. 
    The L\'{e}vy-Khintchine Theorem is applicable to L\'{e}vy processes with measures that have infinite total mass. 
    One example of this is a Gamma process, which has intensity measure on $\Rbb^{+}$ given by $\nu(dy) = ay^{-1}e^{-by}dy$, i.e., on any finite interval of time, the number of jumps which lies in the interval of space $(0,1)$ is infinite. 
    However, we emphasize that we only consider finite-measure L\'{e}vy processes that are formed by taking the affine combination of Gaussian white noise and compound Poisson shot noise processes.
\end{remark}

\subsection{Useful Results}\label{subsec:gronwall}
In this section, we review additional results and inequalities which are used for the proofs of our main theorems in~\sec{stoch_contract}.
First, the well-known \textit{Comparison Lemma} is roughly stated as follows.
Suppose we have an initial-value problem of the form $\{\dot{u} = g(u,t), u(0) = u_0\}$ and corresponding solution $u(t)$. 
Then if we were to consider an analogous problem $\{ D^{+}v \leq g(v,t), v(0) \leq u_0\}$, the solution $v(t)$ satisfies $v(t) \leq u(t)$ for all $t \geq 0$, where $D^{+}$ denotes the right side limit defined as $D^{+}v(t) \triangleq \limsup_{\Delta t\to 0^{+}} (v(t + \Delta t) - v(t))/\Delta t$.
Note that this definition implies that if there exists a function $g(\Delta t,t)$ such that $\abs*{v(t + \Delta t) - v(t)}/\Delta t\leq g(\Delta t,t)$ for all $\Delta t\in(0,a]$ for some $a>0$, and $g_0(t) \triangleq \lim_{\Delta t \to 0+}g(\Delta t,t)$ exists, then $D^{+}v(t) \leq g_0(t)$.
Below, we specialize the Comparison Lemma for the purposes of presenting our main results.

\begin{lemma}[Comparison Lemma for Linear Functions]\label{lem:gronwall_negative_nonconstant}
    Let $y:\Rbb^{+}\to\Rbb^{\geq 0}$ be a piecewise-continuous, nonnegative function which is right-continuous with left limits. Let $\mu > 0$ be a positive constant, and $\theta:\Rbb^{+}\to\Rbb$ be a nonconstant, continuously-differentiable function with $\theta(0)=\zeta > 0$. Fix a value of time $T > 0$, and suppose that there are a finite number $k\in\Nbb$ jump-discontinuities of $y$ in the interval $[0,T)$ which occur at times $t_i, i = 1,\cdots, k$ given by $0 < t_1\leq t_2\leq \cdots\leq t_k<T$.
    Further suppose that there exists a continuous, nonnegative function $h(t)$ which bounds the jumps, i.e., $y(t_i) - y(t_i-) \leq h(t_i)$ for $i = 1,\cdots, k$.
    If the following inequality holds:
    \begin{align}\label{eq:gronwall_negative_nonconstant}
        y(t) - y(0) \leq \theta(t) - \mu\int_{0+}^t y(s-) ds,  \ \ \forall t\in[0,T)
    \end{align}
    Then
    \begin{align}\label{eq:gronwall_negative_nonconstant_result}
        y(t) \leq \int_{0+}^t \frac{d\theta(s)}{ds} e^{-\mu(t-s)}ds + \zeta e^{-\mu t} + y(0)e^{-\mu t}, \ \ \forall t\in[0,T)
    \end{align}
    where the integral from $0+$ to $t$ denotes integration over the interval $(0,t]$.
\end{lemma}

\begin{lemma}[Maximum of Nonnegative Functions]\label{lem:max_nng}
    Suppose $\{\alpha_i\}_{i=1}^k$ and $\{\beta_i\}_{i=1}^k$ are finite, nonnegative, real-valued sequences. 
    Then
    \begin{align}
        \max_{i=1,\cdots,k} \alpha_i\beta_i \leq \max_{i=1,\cdots,k} \alpha_i\max_{i=1,\cdots,k} \beta_i
    \end{align}
\end{lemma}

The proofs to both~\lem{gronwall_negative_nonconstant} and~\lem{max_nng} are straightforward, and have thus been omitted from the paper.
Another useful formula that we employ throughout the paper is \textit{It\^{o}'s Formula}. 
The version of the formula for functions of stochastic processes driven by more general Poisson random measures can be found in various standard references in stochastic processes literature, but for the purposes of this paper, we use the following version of the formula.

\begin{lemma}[It\^{o}'s Formula]\label{lem:ito_formula}
    For functions $G\in \Ccal^{(1,2)}$, i.e., functions $G$ which are once continuously-differentiable in time and twice continuously-differentiable in state:
    \begin{align}\label{eq:ito_shot}
        G(t,\xvect(t)) &= G(0,\xvect_0) + \int_{0+}^{t} \partial_s G(s,\xvect(s-))ds + \sum\limits_{i=1}^n\int_{0+}^{t} \partial_{x_i}G(s,\xvect(s-))dx_i^c(s)\notag\\
        &\hskip1cm + \frac{1}{2}\sum\limits_{i,j=1}^n\int_{0+}^{t} \partial_{x_i}\partial_{x_j} G(s,\xvect(s-)) d[x_i,x_j]^c(s) + \sum\limits_{i=1}^{N(t)} \left( G(T_i,\xvect(T_i)) - G(T_i,\xvect(T_i-))\right)
    \end{align}
    Here, integrals from $0+$ to $t$ indicate an integral over the interval $(0,t]$, $\xvect\in\Rbb^n$, $T_i \in (0, t]$ is the time of the $i$th arrival of $N(t)$, and we use the left-limit notation of Section~\ref{subsec:notation}. 
    Furthermore, $dx_i^c$ represents the continuous part of the SDE for component $x_i$, and $d[x_i, x_j]^c$ represents the continuous part of the quadratic variation between the two SDEs corresponding to components $x_i$ and $x_j$.
\end{lemma}

\begin{definition}[Infinitesimal Generator]\label{def:generator}
    For $G\in\Ccal^{(1,2)}$, the \textit{infinitesimal generator} is defined to be
    \begin{align*}
        \Lcal G = \lim_{t\to 0} \frac{\Ebb_{\xvect_0}\left[ G(t,\xvect(t))\right] - G(0,\xvect_0)}{t}
    \end{align*}
    where $\xvect(t)$ is the trajectory of a given SDE which starts with initial condition $\xvect(0) \triangleq \xvect_0$.
    For the purposes of our paper, this definition extends the standard definition of the infinitesimal generator (see, e.g., (4-12) in Kushner 1967~\cite{kushner67}) to functions $G$ which are also dependent on time $t$. 
\end{definition}

\section{Problem Setup}\label{sec:problem_setup}

\subsection{System Dynamics and Assumptions}\label{subsec:setup}
We consider L\'{e}vy noise systems which can be expressed as SDEs of the following form:
\begin{align}\label{eq:combined_sde}
    d\xvect(t) = f(t, \xvect)dt + \sigma(t,\xvect)dW(t) + \xi(t,\xvect)dN(t)
\end{align}
where 
\begin{itemize}
    \item $f: \Rbb^{+}\times\Rbb^n \to \Rbb^{n}$ is a deterministic function in $\Ccal^{(1,2)}$, i.e., $f$ is once continuously-differentiable in time and twice continuously-differentiable in state.

    \item $\sigma(t,\xvect)dW(t)$ is the additive Gaussian white noise of the system, where $\sigma: \Rbb^{+}\times\Rbb^n\to \Rbb^{n\times d}$, $\sigma \in \Ccal^{(1,2)}$ is the variation of the white noise, and $W: \Rbb^{+}\to\Rbb^d$ is a $d$-dimensional standard Brownian motion process. 

    \item $\xi(t,\xvect)dN(t)$ is a compound Poisson process which enters into the system as an additive disturbance, where $\xi: \Rbb^{+}\times\Rbb^n \to \Rbb^{n}$, $\xi\in\Ccal^{(1,2)}$ describes the jumps that occur, and $N(t)$ is the scalar standard Poisson process with intensity $\lambda>0$.
    The ``derivative'' of the standard Poisson process, written as $dN(t)$, is understood as a function which takes value $1$ if a jump occurs at time $t$, and value $0$ otherwise.
\end{itemize}

\begin{assumption}[Bounded Noise Magnitudes]\label{assum:bdd_noise}
    For the system~\eqn{combined_sde}, there exist constants $\gamma, \eta > 0$ such that $\sup_{t,\xvect}\norm{\sigma(t,\xvect)} \leq \gamma$ and $\sup_{t,\xvect}\norm{\xi(t,\xvect)} \leq \eta$, where the norms are defined in Section~\ref{subsec:notation}.
\end{assumption}

When $\sigma(t, \xvect) \equiv 0$, we have the following \textit{shot noise system}:
\begin{align}\label{eq:shot_sde}
    d\xvect(t) = f(t,\xvect)dt + \xi(t,\xvect)dN(t)
\end{align}
and when $\xi(t,\xvect) \equiv 0$, we have the \textit{white noise system}:
\begin{align}\label{eq:white_sde}
    d\xvect(t) = f(t,\xvect)dt + \sigma(t,\xvect)dW(t)
\end{align}

We choose to use L\'{e}vy noise systems in the form of~\eqn{combined_sde} for its representation simplicity and relevance to real-world system dynamics.
Shot or L\'{e}vy noise systems which can be modeled as~\eqn{shot_sde} or~\eqn{combined_sde} specifically include reinforcement learning-based robust trajectory optimization schemes for robot arm manipulators~\cite{pan15}, stock price fluctuations and impulse control~\cite{oksendal_stoch_ctrl}, and wireless mobile communication networks~\cite{baccelli_book}.

\begin{remark}\label{rmk:generator_ito_same}
    From the formula of the infinitesimal generator, it is easy to see its close relationship with It\^{o}'s formula (\lem{ito_formula}) because it can be used to compute the $\Ebb_{\xvect_0}\left[ G(t,\xvect(t))\right]$ term in~\defin{generator}.
    For example, in the case where~\defin{generator} is applied to a scalar version of the white noise system~\eqn{white_sde} (i.e. $\xvect(t) \triangleq x(t) \in \Rbb$), then the generator is given by $\Lcal G = \partial_t G(t,x(t)) + \partial_xG(t,x(t))f(t,x(t)) + (1/2)\partial_x^2 G(t,x(t)) \sigma^2(t,x(t)) $.
\end{remark}

Throughout this paper, we consider L\'{e}vy noise systems~\eqn{combined_sde} which satisfy the following Lipschitz and bounded-growth conditions to ensure existence and uniqueness of solutions.

\begin{assumption}[Lipschitz and Bounded-Growth]\label{assum:lipschitz_bdd_growth}
    For fixed $T > 0$, consider the restriction of $f,\sigma,\xi$ (defined in~\eqn{combined_sde}) on the time interval $[0,T]$.
    We assume that these functions satisfy the following conditions: 1) $\forall \xvect,\yvect\in\Rbb^n, t\in[0,T]$, $\exists K > 0$ such that $\norm{f(t,\xvect) - f(t,\yvect)} + \norm{\sigma(t,\xvect) - \sigma(t,\yvect)} + \norm{\xi(t,\xvect) - \xi(t,\yvect)} \leq K\norm{\xvect(t) - \yvect(t)}$, and 2) $\forall \xvect\in\Rbb^n, t\in[0,T]$, $\exists C > 0$ such that $\norm{f(t,\xvect)}^2 + \norm{\sigma(t,\xvect)}^2 + \norm{\xi(t,\xvect)}^2 \leq C(1 + \norm{\xvect(t)}^2)$.
\end{assumption}

The conditions for existence and uniqueness of solutions for white noise systems~\eqn{white_sde} are standard; see, e.g., {\O}ksendal 2010~\cite{oksendal_book}.
Using~\assum{lipschitz_bdd_growth} with $\sigma(t,\xvect)\equiv0$, a similar result can be derived in the case of shot noise perturbations.

\begin{lemma}[Existence and Uniqueness for~\eqn{shot_sde}]\label{lem:shot_sde_exist_unique}
    Suppose the functions $f,\xi$ defined in~\eqn{shot_sde} satisfy~\assum{lipschitz_bdd_growth} with $\sigma(t,\xvect)\equiv0$ and some positive constants $C$ and $K$.
    Further, let $\xvect_0 \in \Rbb^n$, $\Ebb[\norm{\xvect_0}] < \infty$, be independent of the shot noise process.
    Then~\eqn{shot_sde} with initial condition $\xvect(0) = \xvect_0$ has a unique solution $\xvect(t)$ adapted to the filtration $\Fcal_t$ generated by $\xvect_0$ and $N(s)$.
    Here, $s\leq t$, $\Ebb_{\xvect_0}[\int_0^T \norm{\xvect(t)}^2 dt] < \infty$, and a solution $\xvect(t)$ of~\eqn{shot_sde} is said to be unique if for any other solution $\yvect(t)$ with initial condition $\yvect(0) = \xvect_0$, we have $\Pbb(\xvect(t) = \yvect(t), \hskip.1cm \forall t>0) = 1$.
\end{lemma}

\begin{remark}
    By the decomposition of L\'{e}vy noise processes into white and shot noise processes via the L\'{e}vy-Khintchine Theorem (see~\remk{levy_khintchine}), the conditions for existence and uniqueness of~\eqn{combined_sde} described in~\assum{lipschitz_bdd_growth} are obtained by combining~\lem{shot_sde_exist_unique} with the Lipschitz and boundedness conditions for white noise systems~\eqn{white_sde}.
    Conditions for more general L\'{e}vy noise systems are similar to~\assum{lipschitz_bdd_growth}, and have been shown in previous literature: see, e.g., Theorem 6.2.3 of Applebaum 2009~\cite{applebaum09_book}.
    There has also been previous work describing conditions for shot noise systems~\eqn{shot_sde} while imposing different, non-Lipschitz conditions on $f$ and $\xi$.
    For instance, Li 2001~\cite{liu01} relaxes the Lipschitz conditions by instead assuming that $f$ and $\xi$ are bounded above by a concave function of the norm difference in trajectories $\norm{\xvect - \yvect}$.
    Alternatively, Kasahara 1991~\cite{kasahara91} presents a result for conditions where $f$ is upper-bounded in norm by a constant, and the bound on $\xi$ depends on the maximum norm bound of the jump. 
    We choose to work with the simple Lipschitz and boundedness conditions of~\assum{lipschitz_bdd_growth} because they are easier to relate to the well-known standard conditions for white noise systems~\eqn{white_sde}.
\end{remark}

\subsection{Incremental Stability}\label{subsec:incremental}
In this section, we set up the \textit{incremental stability} concepts and assumptions required for the presentation of our main results in the subsequent~\sec{stoch_contract}.
For a given stochastic system of the form~\eqn{combined_sde} with initial condition $\xvect(0) = \xvect_0$, suppose we design a Lyapunov-like function $V(t,\xvect)$ such that $\Lcal V \leq -\beta V(t,\xvect(t))$ for some $\beta > 0$ and $\Lcal$ as the infinitesimal generator from~\defin{generator}. 
By Dynkin's formula (see, e.g., (2-9) in Kushner 1967~\cite{kushner67}) and~\lem{gronwall_negative_nonconstant}, we get:
\begin{align}\label{eq:supermartingale_ineq}
    \Ebb_{\xvect_0}[V(t,\xvect(t)] \leq V(0,\xvect_0)e^{-\beta t}
\end{align}
Since $e^{-\beta t} \leq 1$ for all $t\geq 0$,~\eqn{supermartingale_ineq} implies that, under the assumption that $\Ebb_{\xvect_0}[V(t,\xvect(t)] < \infty$ for all $t > 0$, $V(t,\xvect(t))$ is a \textit{supermartingale}.
Using Doob's supermartingale inequality (see, e.g., Chapter 1.7 of Kushner 1967~\cite{kushner67}),~\eqn{supermartingale_ineq} implies:
\begin{align*}
    0 \leq \Pbb\left(\sup_{\tau\leq t<\infty} V(t,\xvect(t)) \geq c\right)
    \leq \frac{1}{c}\Ebb_{\xvect_0}[V(\tau,\xvect(\tau))] \leq \frac{1}{c}V(0,\xvect_0)e^{-\beta \tau}
\end{align*}
for $\tau < t$ and any constant $c > 0$. This further implies \textit{almost-sure stability} of the system since the right side of the inequality tends to $0$ as $\tau\to\infty$. 

The approach we take in the derivation of our main results in~\sec{stoch_contract} is \textit{stochastic contraction theory}.
Our main results to be presented in the following~\sec{stoch_contract} are referred to as the \textit{Shot Noise Stochastic Contraction Theorem} and the \textit{L\'{e}vy Noise Stochastic Contraction Theorem}, and they derive incremental stability conditions for the shot noise system~\eqn{shot_sde} and L\'{e}vy noise system~\eqn{combined_sde}, respectively.
For simplicity, we henceforth refer to each theorem as the \textit{Shot Contraction Theorem} and the \textit{L\'{e}vy Contraction Theorem}, respectively.
In the following discussion, we formalize the notion of incremental stability and contraction in both the deterministic and stochastic sense.

\begin{definition}[Incremental Exponential Stability]\label{def:increm_stable}
    For any nonlinear function $f\in \Ccal^{(1,2)}$, the deterministic system $d\xvect = f(t, \xvect)dt$ is said to be \textit{incrementally (globally exponentially) stable} if there exist constants $\kappa,\alpha > 0$ such that
    \begin{align}\label{eq:increm_stable}
        \norm{\xvect_2(t) - \xvect_1(t)} \leq \kappa\norm{\xvect_{2,0} - \xvect_{1,0}}e^{-\alpha t}
    \end{align}
    for any vector norm $\norm{\cdot}$, and all $t \geq 0$. The trajectories $\xvect_1(t)$ and $\xvect_2(t)$ are solutions of the system $d\xvect = f(t, \xvect)dt$ with respective initial conditions $\xvect_{1,0}$ and $\xvect_{2,0}\neq\xvect_{1,0}$. 
\end{definition}

Following the notation from~\defin{increm_stable}, we denote $\delta \xvect\in\Rbb^n$ to be the \textit{infinitesimal displacement length} between $\xvect_1(t)$ and $\xvect_2(t)$ over a fixed infinitesimal interval of time. Formally, the infinitesimal displacement length is represented as a path integral:
\begin{align}\label{eq:mu_param}
    \norm{\xvect_2(t) - \xvect_1(t)} \leq \int_{\xvect_1}^{\xvect_2} \norm{\delta\xvect(t)}
\end{align}

The evolution of the infinitesimal displacement over time can be approximated by the dynamics
\begin{align}\label{eq:Fdef}
    d\delta\xvect = F\delta\xvect dt
\end{align}
where $F \triangleq \nabla_{\xvect} f(t,\xvect)$ is the \textit{Jacobian} of the system.
These dynamics, associated with the state $\delta\xvect$, are commonly referred to as the \textit{virtual dynamics}.

Similar to the indirect and direct Lyapunov methods of testing Lyapunov stability, there is a test to determine incremental stability of a system without needing the literal~\defin{increm_stable}.
Oftentimes, performing a differential coordinate transform from $\delta\xvect$ to $\delta\zvect \triangleq \Theta(t,\xvect)\delta\xvect$, where $\Theta(t,\xvect) \in \Rbb^{n\times n}$ is a smooth invertible square matrix, makes it easier to verify the conditions of this test.
The new virtual dynamics under this coordinate transform become
\begin{align}\label{eq:virtual_dynamics_trans}
    d\delta\zvect = F_g\delta\zvect dt
\end{align}
where
\begin{align}\label{eq:Fdef_trans}
    F_g \triangleq (\dot{\Theta}(t,\xvect) + \Theta(t,\xvect) F)\Theta^{-1}(t,\xvect)
\end{align}
is the \textit{generalized Jacobian} of the system, and the dot notation is defined in Section~\ref{subsec:notation}.

An equivalent way to say that a system $d\xvect(t) = f(t,\xvect)dt$ is incrementally stable in the sense of~\defin{increm_stable} is to say that it is \textit{contracting} with some rate $\alpha > 0$. 
Similar to~\eqn{virtual_dynamics_trans}, we can extend the notion of contraction to more general metrics: a system is incrementally stable if it is contracting with respect to a uniformly positive definite metric $M(t,\xvect) \triangleq \Theta(t,\xvect)^T\Theta(t,\xvect)$ and convergence rate $\alpha$. 
For most practical applications, we are able to make the following assumption on $M(t,\xvect)$.
    
\begin{assumption}[Bounded Metric]\label{assum:metric_bounds}
    The metric $M(t,\xvect)$ described in the setup above is bounded in both arguments $\xvect$ and $t$ from above and below, and its first and second derivatives with respect to the $\xvect$ argument are also bounded from above. We thus define the following constants
    \begin{align}\label{eq:metric_bounds}
        &\underline{m} = \inf_{t,\xvect} \lambda_{\text{min}}(M(t,\xvect)), \qquad \overline{m} = \sup_{t,\xvect} \lambda_{\text{max}}(M(t,\xvect))\\
        &m' = \sup_{t,\xvect, i, j} \norm{(\partial_x M(t,\xvect))_{i,j}},
        \qquad m'' = \sup_{t,\xvect, i, j} \norm{(\partial_x^2 M(t,\xvect))_{i,j}}\notag
    \end{align}
\end{assumption}

The inequality \eqn{increm_stable} is obtained for the special case where $M(t,\xvect) = I_n$, the $n$-dimensional identity matrix. For general deterministic system dynamics $d\xvect(t) = f(t,\xvect)dt$, the criterion for testing incremental stability is stated in the theorem below.

\begin{theorem}[Basic Contraction]\label{thm:basic_contract}
    Consider the deterministic dynamics $d\xvect = f(t,\xvect)dt$. If there exists a uniformly positive definite metric $M(t,\xvect)$ and $\alpha > 0$ such that the following condition is satisfied:
    \begin{align}\label{eq:norm_squared_cond}
        F^T M(t,\xvect) + M(t,\xvect)F + \dot{M}(t,\xvect) \leq -2\alpha M(t,\xvect)
    \end{align}
    then the system is contracting. Moreover, in relation to~\eqn{Fdef_trans}, we have $\lambda_{\text{max}}((F_g+F_g^T)/2) \leq -\alpha$.
\end{theorem}

For deterministic systems, incremental stability has been established as a concept of convergence between different solution trajectories with different initial conditions~\cite{lohmiller98,aminzare14}. However, in the stochastic setting, the difference between trajectories also arises from using different noise processes.
For this reason, we require a change in notation from deterministic incremental stability analysis.
The infinitesimal displacement length now considers the difference between a solution trajectory $\xvect(t)$ of a stochastic system with one noise sample path and a solution trajectory $\yvect(t)$ of a stochastic system with another noise sample path.
This can be viewed as a comparison between solution trajectories coming from distinct systems; this is different from deterministic incremental stability, which compares of two solution trajectories from the same system.
To make this distinction clear, we use the notation $\delta\qvect$ in place of $\delta\xvect$, and the path integral~\eqn{mu_param} is now written instead with a parametrization $\mu\in[0,1]$:
\begin{align}\label{eq:mu_param_stoch}
    \norm{\yvect(t) - \xvect(t)} \leq \int_{\xvect}^{\yvect} \norm{\delta\qvect(t)} = \int_0^1\norm{ \partial_{\mu} \qvect(\mu,t)} d\mu
\end{align}

The work of Pham 2009~\cite{pham09} considered stochastic incremental stability for the specific case of additive Gaussian white noise perturbations, and Dani 2015~\cite{dani15} extended this theory to more general state-dependent metrics. 
Both works compared between two noise-perturbed trajectories -- $\xvect(t)$, solution to~\eqn{white_sde} with white noise $\sigma_1(t,\xvect)dW_1(t)$, and $\yvect(t)$, solution to~\eqn{white_sde} with white noise $\sigma_2(t,\xvect)dW_2(t)$.
However, in this paper, we compare one noise-perturbed trajectory $\xvect(t)$ against a trajectory $\yvect(t)$ of the nominal, deterministic system $d\yvect = f(t,\yvect)dt$.
This allows for a direct combination of the white noise result with the Shot Contraction Theorem to establish the L\'{e}vy Contraction Theorem.
To that end, we consider a parametrization of a new state $\qvect(\mu,t)\in\Rbb^n$, with $\mu\in[0,1]$, such that:
\begin{gather}\label{eq:levy_noise_mu}
    \qvect(\mu = 0, t) = \xvect(t),\ \qvect(\mu = 1, t) = \yvect(t), \quad \sigma_{\mu = 0}(t,\qvect) = \sigma(t,\xvect),\ \sigma_{\mu = 1}(t,\qvect) = 0, \quad \xi_{\mu = 0}(t,\qvect) = \xi(t,\xvect), \ \xi_{\mu = 1}(t,\qvect) = 0
\end{gather}
where $\xvect(t)$ and $\yvect(t)$ are solution trajectories of, respectively:
\begin{subequations}
    \begin{align}
        d\xvect(t) &= f(t,\xvect)dt + \sigma(t,\xvect)dW(t) + \xi(t, \xvect)dN(t), \quad \xvect(0) = \xvect_0\label{eq:perturbed_system}\\
        d\yvect(t) &= f(t,\yvect)dt, \quad \yvect(0) = \yvect_0\label{eq:nominal_system}
    \end{align}
\end{subequations}
This parametrization allows us to construct a \textit{virtual system} with state $\qvect(\mu,t)$, written as
\begin{align}\label{eq:virtual_system}
    d\qvect(\mu,t) = f(t,\qvect(\mu,t))dt + \sigma_{\mu}(t,\qvect(\mu,t))dW(t) + \xi_{\mu}(t,\qvect(\mu,t))dN(t)
\end{align}
The virtual dynamics become
\begin{align}\label{eq:virtual_dynamics}
    d\delta\qvect(t) = F\delta\qvect(t) dt + \delta\sigma_{\mu}dW(t) + \delta\xi_{\mu}dN(t)
\end{align}
where $F$ is the Jacobian defined in~\eqn{Fdef} and
\begin{align}\label{eq:virtual_notation}
    \delta\sigma_{\mu}(t,\qvect) \triangleq \left[\nabla_{\qvect}\sigma_{\mu,1}(t,\qvect)\delta\qvect(t), \cdots, \nabla_{\qvect}\sigma_{\mu,d}(t,\qvect)\delta\qvect(t) \right]\in\Rbb^{n\times d},\quad
    \delta\xi_{\mu}(t,\qvect) \triangleq \nabla_{\qvect}\xi_{\mu}(t,\qvect)\delta\qvect(t)\in\Rbb^{n}
\end{align}
where $\sigma_{\mu,i}$ is the $i$th column of $\sigma_{\mu}$.

For the white noise system~\eqn{white_sde}, the perturbed system and its nominal dynamics are parametrized such that the virtual system and virtual dynamics are established as in~\eqn{virtual_system} and~\eqn{virtual_dynamics} without the shot noise terms $\xi_{\mu}(t,\qvect)$, $\delta\xi{\mu}$, and $dN(t)$.
Stochastic contraction is defined in Definition 2 of Pham 2009~\cite{pham09}, but is only applicable to white noise systems~\eqn{white_sde}.
For the virtual system~\eqn{virtual_system}, we create a more general definition of stochastic contraction.

\begin{definition}[Stochastically Contracting]\label{def:stoch_contract}
    The system~\eqn{combined_sde} is said to be \textit{stochastically contracting} if:
    \begin{enumerate}
        \item the nominal, unperturbed system $d\xvect = f(t,\xvect)dt$ is contracting with some differential coordinate transform $\Theta(t,\xvect)$ and convergence rate $\alpha$, i.e.,~\eqn{norm_squared_cond} is satisfied.
        
        \item there exists a Lyapunov-like function $V(t,\xvect)$ such that
        \begin{align}\label{eq:supermartingale_ineq_error}
            \Ebb[V(t,\qvect(t))] \leq V(0,\qvect_0)e^{-\beta t} + \kappa(t)
        \end{align}
        for some contraction rate $\beta > 0$ and bounded function $\kappa:\Rbb^{+}\to\Rbb^{+}$, and initial condition $\qvect_0 \in\Rbb^n$.
    \end{enumerate}
\end{definition}

\begin{remark}
    The equation~\eqn{supermartingale_ineq_error} is a version of~\eqn{supermartingale_ineq} with a nonzero steady-state error bound $\kappa(t)$. 
    This is because for stochastic systems, convergence to an equilibrium often does not occur with perfectly zero error due to trajectories arising from different noise sample paths.
    Moreover, for the impulsive shot noise in~\eqn{combined_sde} and~\eqn{shot_sde}, almost-sure convergence is difficult to demonstrate.
    We show this in~\sec{stoch_contract}, where the bound is instead derived as a probabilistic guarantee, conditioning on a finite number of jumps within a fixed interval of time.
\end{remark}

Demonstrating stochastic incremental stability for stochastic systems perturbed by some class of noise processes involves rewriting~\eqn{supermartingale_ineq_error} and deriving specific forms of $\beta$ and $\kappa(t)$ based on the parameters of the stochastic system. 
One common choice of Lyapunov function is the metric-weighted norm-squared difference between solution trajectories with distinct initial conditions and noise sample paths:
\begin{align}\label{eq:sde_lyap}
    V(t, \qvect(\mu,t), \delta\qvect(t)) = \int_0^1 \partial_{\mu}\qvect^T(\mu,t)M(t,\qvect(\mu,t))\partial_{\mu}\qvect(\mu,t) d\mu
\end{align}
Here, $M(t,\qvect) \triangleq \Theta^T\Theta(t,\qvect)$ is the contraction metric described before; the parametrization over $\mu$ is such that~\eqn{virtual_system} and~\eqn{virtual_dynamics} hold and $V(0,\qvect_0,\delta\qvect_0) = \norm{\yvect_0 - \xvect_0}$.
Similar to the direct method of Lyapunov, we analyze the behavior of the system by analyzing the derivative of the Lyapunov-like function $V(t,\qvect,\delta\qvect)$ along trajectories of the virtual system~\eqn{virtual_system}. 

\begin{remark}\label{rmk:finite_horz}
    Another significant distinction between our stochastic incremental stability setup and previous versions is that we derive an error bound over a fixed interval of time $[s,t]$ for any $s < t$ instead of necessarily fixing $s=0$ and including the initial state.
    This allows us to interpret our stability theorems as a measure of how far the perturbed trajectory will deviate from the nominal within a local horizon of time, which allows for the design of controllers and observers which are online and adaptive.
\end{remark}

\begin{theorem}[White Noise Stochastic Contraction Theorem]\label{thm:white_contract}
    Suppose that the perturbed system~\eqn{white_sde} is stochastically contracting in the sense of~\defin{stoch_contract} under a differential coordinate transform $\Theta(t,\xvect)$ which satisfies~\assum{metric_bounds}.
    Then, for a fixed interval of time $[s,t]$ for $s<t$,~\eqn{supermartingale_ineq_error} can be written explicitly as:
    \begin{align}\label{eq:white_contract_bound_2}
        \Ebb[\norm{\yvect(t) - \xvect(t)}^2] \leq \frac{1}{\underline{m}} \Ebb\left[\norm{\yvect(s) - \xvect(s)}^2\right]e^{-\beta_w (t-s)} + \frac{\kappa_w(\beta_w,s,t)}{\underline{m}\beta_w}
    \end{align}
    where
    \begin{subequations}\label{eq:white_params}
        \begin{align}
            \beta_w &= 2\alpha - \frac{\gamma^2}{\underline{m}}\left(m' + \frac{m''}{2}\right)\\
            \kappa_w(\beta_w,s,t) &= \gamma^2(m'+\overline{m})\left( 1 - e^{-\beta_w (t - s)}\right)
        \end{align}
    \end{subequations}  
    and $\gamma$ is the bound defined in~\assum{bdd_noise}, $\underline{m}, \overline{m}, m', m''$ are the constants defined in~\assum{metric_bounds}, and $\alpha$ is the deterministic contraction rate from~\thm{basic_contract}.
\end{theorem}

Dani 2015~\cite{dani15} demonstrates an application of white noise incremental stability to the problem of model-based nonlinear observer design.
Thus, extending~\thm{white_contract} to account for non-Gaussian noise gives us a potential way to design model-based observers and controllers for systems perturbed by non-Gaussian noise. 
With this motive, the next section presents our main results, the Shot Contraction Theorem and the L\'{e}vy Contraction Theorem, which are incremental stability theorems for the shot noise system~\eqn{shot_sde} and L\'{e}vy noise system~\eqn{combined_sde}, respectively.


\section{Stochastic~Contraction~Theorems}\label{sec:stoch_contract}
Given our setup established in Section~\ref{subsec:incremental}, we present the Shot Contraction Theorem for shot noise systems~\eqn{shot_sde} and the L\'{e}vy Contraction Theorem for L\'{e}vy noise systems~\eqn{combined_sde}, and use them to conclude incremental stability properties for each respective system.
The proofs to both theorems follow similar approaches: we analyze the behavior of the system by analyzing the derivative of the Lyapunov-like function $V(t,\qvect,\delta\qvect)$ along trajectories of the virtual system~\eqn{virtual_system}. 
This requires us to use the infinitesimal generator from~\defin{generator}, which can be thought of as the stochastic analogue to the differentiation operator used in the deterministic case. 
For the shot and L\'{e}vy noise systems, we invoke It\^{o}'s formula (\lem{ito_formula}) instead of using the infinitesimal generator by virtue of the relationship described in~\remk{generator_ito_same}.
Throughout this section, we shorten the notation such that $V(t,\qvect,\delta\qvect)$ is understood to mean $V(t,\qvect(\mu,t),\delta\qvect(t))$, i.e., the spatial arguments $\qvect$ and $\delta\qvect$ are evaluated at the time argument $t$.

\subsection{Main Results}\label{subsec:stoch_contract_thms}
We begin with the Shot Contraction Theorem for shot noise systems~\eqn{shot_sde}.
We compare a trajectory $\xvect(t)$ of the shot noise system~\eqn{shot_sde} against a trajectory $\yvect(t)$ of the nominal system~\eqn{nominal_system}.
We define the parameter $\mu\in[0,1]$ such that the virtual system and virtual dynamics are established as in~\eqn{virtual_system} and~\eqn{virtual_dynamics} without the white noise terms $\sigma_{\mu}(t,\qvect)$, $\delta\sigma_{\mu}$, and $dW(t)$.

\begin{assumption}[Bounded Differences of Lyapunov-Like Function]\label{assum:theorem_assumptions}
    Consider the shot noise system~\eqn{shot_sde} and corresponding Lyapunov-like function~\eqn{sde_lyap}.
    For any fixed $t > 0$, there exists a deterministic, locally-bounded, continuously-differentiable function $h: \Rbb^{+} \to \Rbb^{+}$ such that
    \begin{align}\label{eq:time_var_bound}
        \Ebb \left[ \sup_{r\in[0,t]} V(r,\qvect, \delta\qvect) - V(r-,\qvect, \delta\qvect)\right] \leq h(t)
    \end{align}
\end{assumption}

For the remainder of this section, the expectation operator $\Ebb[\cdot]$ is understood to be taken over all sources of randomness in the argument. 
For instance, in~\assum{theorem_assumptions}, $\Ebb[\cdot]$ is taken with respect to the initial condition distribution $p(\qvect_0)$, the random function $\xi$ describing the jump distribution, and the standard Poisson process $N(t)$.

\begin{remark}\label{rmk:abstract_c}
    Because the Lyapunov-like function $V(t,\qvect,\delta\qvect)$ takes in arguments $\qvect$ and $\delta\qvect$ which depend on the shot noise process of~\eqn{shot_sde},
    more information is needed about $f(t,\xvect)$ and $\xi(t,\xvect)$ in order to design a metric $M(t,\qvect)$ for $V(t,\qvect,\delta\qvect)$.
    In Section~\ref{subsec:shot_contract_linear}, we make the abstract function $h(t)$ in~\assum{theorem_assumptions} more concrete by specializing~\eqn{shot_sde} to linear time-varying (LTV) systems where $f(t,\xvect) \triangleq A(t)\xvect$ and jumps $\xi(t,\xvect)\equiv\xi(t)$ which are independent of the state $\xvect$. 
    We will see that the expression of $h(t)$ in terms of system parameters depends on the form of the solution trajectory $\xvect(t)$ of~\eqn{shot_sde}, which is easy to obtain for LTV systems.
\end{remark}

\begin{remark}\label{rmk:jumps_boundedness}
    The existence of a $h(t)$ in~\eqn{time_var_bound} is roughly justified using the following argument.
    For each time $t \geq 0$, note that we can simplify the difference in~\eqn{time_var_bound} as follows:
    \begin{align}\label{eq:cz0_exist}
        &V(t,\qvect,\delta\qvect) - V(t-,\qvect,\delta\qvect)\notag\\
        &\hskip2cm = \int_0^1 \left[ \partial_{\mu}\qvect(\mu,t)^TM(t,\qvect(\mu,t))\partial_{\mu}\qvect(\mu,t) - \partial_{\mu}\qvect(\mu,t-)^TM(t,\qvect(\mu,t))\partial_{\mu}\qvect(\mu,t-)\right] d\mu\notag\\
        &\hskip2cm\leq \int_0^1 \left(\overline{m}\norm{\partial_{\mu}\qvect(\mu,t)}^2 - \underline{m}\norm{\partial_{\mu}\qvect(\mu,t-)}^2\right)d\mu
    \end{align}
    where the last inequality follows from~\assum{metric_bounds}.
    By continuity of the nominal system, $\yvect(t) = \yvect(t-)$. Hence, the difference~\eqn{cz0_exist} is nonzero only when there exists a jump of the shot noise process at time $t$.
    By~\assum{bdd_noise}, we know that the jumps are bounded by a constant $\eta$, and so the difference~\eqn{cz0_exist} is also bounded at each fixed time $t$.
\end{remark}

\begin{definition}[Condition on the Number of Jumps]\label{def:cond_k}
    Let $N(t)$ be the standard Poisson process driving the shot noise process behind systems of the form~\eqn{shot_sde} or~\eqn{combined_sde}.
    For fixed values of time $0\leq s\leq t$, define $\Ebb_k[\cdot] \triangleq \Ebb[\cdot|N(t) - N(s) = k]$ to be the expectation operator $\Ebb[\cdot]$ conditioned on the occurrence of $k\in\Nbb$ jumps within the fixed interval of time $[s,t]$.
    By the stationarity property of Poisson processes, the event is equivalent to the event that $N(t-s) = k$, which occurs with probability \begin{align}\label{eq:poisson_prob}
        p_k(t-s) \triangleq \Pbb(N(t) - N(s) = k) = e^{-\lambda (t-s)} \frac{(\lambda (t-s))^k}{k!}
    \end{align}
    and we have the relationship $\Ebb[\cdot] = \sum_{k=0}^{\infty}p_k\Ebb_k[\cdot]$.
\end{definition}

\begin{theorem}[Shot Noise Stochastic Contraction Theorem]\label{thm:shot_contract}
    Suppose that the shot noise system~\eqn{shot_sde} is perturbed by noise processes which satisfy~\assum{bdd_noise}, and is stochastically contracting in the sense of~\defin{stoch_contract} under a differential coordinate transform $\Theta(t,\qvect)$.
    Further suppose the metric $M(t,\qvect)$ constructed from $\Theta(t,\qvect)$ satisfies~\eqn{metric_bounds}, and is such that the Lyapunov-like function~\eqn{sde_lyap} satisfies~\assum{theorem_assumptions}. 
    If for a fixed interval of time $[s,t]$ for $0 \leq s < t$, $k\in\Nbb$ jumps occur with probability $p_k(t-s)$ given by~\eqn{poisson_prob}, then~\eqn{supermartingale_ineq_error} can be written explicitly as:
    \begin{align}\label{eq:shot_lyap_bound}
        \Ebb_k[\norm{\yvect(t) - \xvect(t)}^2] \leq \frac{1}{\underline{m}} \Ebb_k[\norm{\yvect(s) - \xvect(s)}^2]e^{-\beta_s (t - s)} + \frac{\kappa_s(\beta_s,s,t)}{\underline{m}}
    \end{align}
    where
    \begin{subequations}\label{eq:shot_params}
        \begin{align}
            \beta_s &\triangleq 2\alpha\label{eq:beta_shot}\\
            \kappa_s(\beta_s,s,t) &\triangleq 
            k\int_{s+}^t \frac{dh(\tau)}{d\tau} e^{-\beta_s(t-\tau)}d\tau + kh(s) e^{-\beta_s(t-s)}\label{eq:kappa_shot}
        \end{align}
    \end{subequations}
    and $\alpha > 0$ is the deterministic contraction rate from~\eqn{norm_squared_cond},
    $\underline{m}$ is defined in~\eqn{metric_bounds}, and the function $h$ is defined in~\assum{theorem_assumptions}.
\end{theorem}

\begin{proof}[Proof of~\thm{shot_contract}]
Apply~\lem{ito_formula} to~\eqn{sde_lyap}, and apply $\Ebb_{k}$ across the resulting equation to condition on the number of jumps being $N(t) - N(s) = k$. We get:
\begin{subequations}\label{eq:shot_contract_eq_0_5}
    \begin{align}
        \Ebb_{k}\left[V(t,\qvect,\delta\qvect)\right] - \Ebb_{k}\left[V(s,\qvect,\delta\qvect)\right] &= \Ebb_{k}\left[\int_{s+}^t \partial_{\tau} V(\tau,\qvect,\delta\qvect)d\tau\right]\label{eq:shot_contract_eq_0_5_det1}\\
        &+ \sum\limits_{i=1}^n\Ebb_{k}\bigg[\int_{s+}^t \left( \partial_{q_i}V(\tau,\qvect,\delta\qvect) f_i(\tau,\qvect) + \partial_{\delta q_i}V(\tau,\qvect,\delta\qvect) \left( F\delta\qvect\right)_i\right) d\tau\bigg]\label{eq:shot_contract_eq_0_5_det2}\\
        &+ \Ebb_k\left[ \sum\limits_{i= N(s)+1}^{N(t)} \left( V(T_i,\qvect,\delta\qvect) - V(T_i-,\qvect,\delta\qvect)\right)\right]\label{eq:shot_contract_eq_0_5_shot}
    \end{align}
\end{subequations}
where $F$ is the Jacobian from~\eqn{Fdef}, $T_i$ is the time of the $i$th arrival in the Poisson process $N(t)$ driving the shot noise system~\eqn{shot_sde}, and we use the left-limit notation of Section~\ref{subsec:notation}.
Similar to the argument of~\remk{jumps_boundedness}, note that each term of the sum~\eqn{shot_contract_eq_0_5_shot} is nonzero only if there is a jump at time $s$, where $s\leq t$. 
Furthermore, the terms of~\eqn{ito_shot} which correspond to the continuous part of the quadratic variation are zero for dynamics~\eqn{shot_sde}.

Recall that the Lyapunov-like function~\eqn{sde_lyap} is twice continuously-differentiable with respect to its arguments $\qvect$ and $\delta\qvect$.
This means there is a jump-discontinuity in $V$ only if there is a jump discontinuity in $\qvect$ or $\delta\qvect$. 
But by the relationship between~\eqn{virtual_system} and~\eqn{virtual_dynamics}, $\qvect$ and $\delta\qvect$ experience jumps at the same times.
Hence, the number of jumps experienced by $V$ in a fixed interval of time $[s,t]$ is equal to the number of jumps experienced by the trajectory $\qvect$ in $[s,t]$. 

A bound on~\eqn{shot_contract_eq_0_5_det1} and~\eqn{shot_contract_eq_0_5_det2} is derived from~\thm{basic_contract}.
We bound~\eqn{shot_contract_eq_0_5_shot} in the following way:
\begin{subequations}\label{eq:shot_contract_eq_0_25}
    \begin{align}
        \Ebb_k\left[ \sum\limits_{i= N(s)+1}^{N(t)} \left( V(T_i,\qvect,\delta\qvect) - V(T_i-,\qvect,\delta\qvect)\right)\right] &= \Ebb_k\left[\sum\limits_{i=1}^{k} \left(V(T_i,\qvect,\delta\qvect) - V(T_i-,\qvect,\delta\qvect)\right)\right]\label{eq:shot_contract_eq_0_25_1}\\
        &\leq k\Ebb_k\left[\max_{i\in\{1,\cdots,k\}} \left(V(T_i,\qvect,\delta\qvect) - V(T_i-,\qvect,\delta\qvect\right)\right]\label{eq:shot_contract_eq_0_25_2}\\
        &\leq kh(t)\label{eq:shot_contract_eq_0_25_3}
    \end{align}
\end{subequations}
where $h(t)$ is defined in~\assum{theorem_assumptions}.
The inequality~\eqn{shot_contract_eq_0_25_3} comes from~\eqn{time_var_bound} and the fact that $T_i\in[s,t]$ for all $i=1,\cdots,k$.
In~\eqn{shot_contract_eq_0_25_1}, we abuse the notation for the subscript $i$ in $T_i$ for both sums which range over $i = N(s) + 1$ to $N(t)$ and sums which range over $i=1$ to $k$.
This is done for the sake of simplicity, with the understanding that~\eqn{shot_contract_eq_0_25_1} arises because we conditioned on $N(t) - N(s) = k$.

In combination, we get:
\begin{align}\label{eq:shot_contract_eq_5}
    \Ebb_{k}\left[ V(t,\qvect,\delta\qvect) \right] - \Ebb_{k}\left[V(s,\qvect,\delta\qvect)\right] &\leq -2\alpha\int_{s+}^{t} \Ebb_{k}[V(\tau,\qvect,\delta\qvect)] d\tau + kh(t)
\end{align}
where $\alpha$ is the contraction rate of the nominal system. 
Applying~\lem{gronwall_negative_nonconstant} with $y(t) \triangleq \Ebb_k[V(t,\qvect,\delta\qvect)]$, $\zeta = kh(s)$, $\mu \triangleq 2\alpha$, and $\theta(t) \triangleq kh(t)$ turns~\eqn{shot_contract_eq_5} into
\begin{align}\label{eq:shot_contract_eq_6}
    \Ebb_k[V(t,\qvect,\delta\qvect)] \leq \Ebb_k[V(s,\qvect,\delta\qvect)]e^{-\beta_s (t-s)} + \kappa_s(\beta_s,s,t)
\end{align}
where $\beta_s$ and $\kappa_s(\beta_s,s,t)$ are defined in~\eqn{beta_shot} and~\eqn{kappa_shot}, respectively. Note that by~\assum{metric_bounds},~\eqn{mu_param_stoch}, and Cauchy-Schwarz:
\begin{align}\label{eq:shot_contract_eq_7}
    \underline{m}\Ebb_k\left[\norm{\yvect(t) - \xvect(t)}^2\right] \leq \underline{m}\Ebb_k\left[\int_0^1\norm{ \partial_{\mu} \qvect(\mu,t)}^2 d\mu\right] \leq \Ebb_k[V(t,\qvect,\delta\qvect)]
\end{align}
We use~\eqn{shot_contract_eq_7} to write~\eqn{shot_contract_eq_6} as an inequality on the norm mean-squared error between the two trajectories $\xvect$ and $\yvect$.
Because the condition that $N(t) - N(s) = k$ occurs with probability $p_k(t-s)$ given by~\eqn{poisson_prob}, we obtain our desired bound~\eqn{shot_lyap_bound}.
\end{proof}

Both the shot noise system and the nominal system have the same contraction rate. 
This is because the shot noise system behaves exactly as the nominal system in between consecutive jumps.
The displacements of the shot noise trajectory incurred by the jumps of the shot noise process can each be thought of as a reset to a different initial condition from which the system evolves nominally.

\begin{remark}\label{rmk:white_vs_shot_bounds}
    In conjunction with~\remk{finite_horz}, we note two important differences between white noise incremental stability~\thm{white_contract} and shot noise incremental stability~\thm{shot_contract}.
    The first difference is that taking $t\to\infty$ in the inequality of~\thm{white_contract} yields a bound which can be interpreted as the steady-state error ball that solution trajectories are guaranteed to converge towards.
    In contrast,~\thm{shot_contract} is more comparable to finite-time stability theory, described in Chapter III of Kushner 1967~\cite{kushner67}.
    The second difference is that, ue to the impulsive, large-norm jumps of the noise process, the mean-squared error bound for the cases of shot noise systems are provided with a specific probability of satisfaction; this probability is dependent upon the number of jumps incurred by the noise process.
    Depending on the application, such criteria may be viewed as being weaker than the traditional steady-state, mean-squared sense of convergence.
    However, for other applications which deal with online implementations of controller/observer synthesis, this probabilistic guarantee potentially allows for better time-varying, adaptive design.
    For instance, the probability of~\eqn{poisson_prob} can be used as a measure of expectation that $k$ jumps will arise in a fixed, future horizon of time; given this event, the bound of~\thm{shot_contract} provides a guarantee on the mean-squared deviation of the perturbed trajectory away from the nominal.
\end{remark}

\begin{remark}\label{rmk:tradeoff}
    The error ball~\eqn{kappa_shot} depends on two types of parameters: 1) fixed, inherent parameters which come from the system dynamics/noise process, and 2) design parameters which can be tuned to vary the stability bounds.
    Among the inherent parameters, we can make the following observation about the intensity $\lambda$ of the shot noise process.
    An increasing $\lambda$ is indicative of a more rapid accumulation of jumps, which implies larger deviations of the perturbed trajectory away from the nominal over shorter horizons of time.
    This can be seen by the error ball $\kappa_s$ in~\eqn{kappa_shot} being directly proportional to the number of jumps $k$, and we demonstrate this relationship numerically using the specific 2D nonlinear system in Section~\ref{subsec:2D_nonlinear}.
    The effects of other parameters vary based on the type of function $f$ which describes the unperturbed dynamics.
    Hence, further discussion is deferred to Section~\ref{subsec:shot_contract_linear}, which provides further insights specifically for linear time-varying (LTV) forms of~\eqn{shot_sde}.
\end{remark}

Now we are ready to present the L\'{e}vy Contraction Theorem for the L\'{e}vy noise system~\eqn{combined_sde}.
We show that the resulting condition is a combination of the conditions for white noise (\thm{white_contract}) and the shot noise (\thm{shot_contract}).
Consider two trajectories of a system -- $\xvect(t)$ a solution of~\eqn{perturbed_system} and $\yvect(t)$ a solution of the nominal system~\eqn{nominal_system}.
We define the parameter $\mu\in[0,1]$ which yields the parametrization~\eqn{levy_noise_mu}, virtual system~\eqn{virtual_system}, and virtual dynamics~\eqn{virtual_dynamics}.
Analogous to the white noise parameters $\beta_w, \kappa_w$ from~\eqn{white_params} and the shot noise parameters $\beta_s, \kappa_s$ from~\eqn{shot_params}, denote $\beta_{\ell},\kappa_{\ell}$ to be the contraction rate and steady-state error bound, respectively, for the L\'{e}vy noise system. 

\begin{theorem}[L\'{e}vy Noise Stochastic Contraction Theorem]\label{thm:levy_contract}
    Suppose that the L\'{e}vy noise system~\eqn{combined_sde} is perturbed by noise processes which satisfy~\assum{bdd_noise}, and is stochastically contracting in the sense of~\defin{stoch_contract} under a differential coordinate transform $\Theta(t,\qvect)$.
    Further suppose the metric $M(t,\qvect)$ constructed from $\Theta(t,\qvect)$ satisfies~\eqn{metric_bounds}, and is such that the Lyapunov-like function~\eqn{sde_lyap} satisfies~\assum{theorem_assumptions}.
    If for a fixed interval of time $[s,t]$ for $0 \leq s < t$, $k\in\Nbb$ jumps occur with probability $p_k(t-s)$ given by~\eqn{poisson_prob}, then~\eqn{supermartingale_ineq_error} can be written explicitly as:
    \begin{align}\label{eq:levy_lyap_bound}
        \Ebb_k[\norm{\yvect(t) - \xvect(t)}^2] \leq \frac{1}{\underline{m}} \Ebb_k\left[\norm{\yvect(s) - \xvect(s)}^2\right]e^{-\beta_{\ell} (t-s)} + \frac{\kappa_{\ell}(\beta_{\ell},s,t)}{\underline{m}}
    \end{align}
    where
    \begin{subequations}\label{eq:levy_params}
        \begin{align}
            \beta_{\ell} &\triangleq 2\alpha - \frac{\gamma^2}{\underline{m}}\left( m' + \frac{m''}{2} \right) = \beta_w\label{eq:beta_levy}\\
            \kappa_{\ell}(\beta_{\ell},s,t) &\triangleq k\int_{s+}^t \frac{dh(\tau)}{d\tau} e^{-\beta_{\ell}(t-\tau)}d\tau + kh(s)e^{-\beta_{\ell}(t-s)} + \frac{\gamma^2}{\beta_{\ell}}\left( m'+\overline{m}\right)\left(1 - e^{-\beta_{\ell}(t-s)} \right)\notag\\
            &= \kappa_s(\beta_{\ell},s,t) + \frac{1}{\beta_{\ell}}\kappa_w(\beta_{\ell},s,t) = \kappa_s(\beta_{w},s,t) + \frac{1}{\beta_{w}}\kappa_w(\beta_{w},s,t)\label{eq:kappa_levy}
        \end{align}
    \end{subequations}
    and $\alpha$ is the deterministic contraction rate from~\eqn{norm_squared_cond}, $\gamma$ is the norm bound on the variation of the white noise process from~\assum{bdd_noise}, and $\underline{m}, \overline{m}, m', m''$ are defined in~\eqn{metric_bounds}. The function $h$ is defined in~\assum{theorem_assumptions}.
\end{theorem}

\begin{proof}[Proof of~\thm{levy_contract}]
    Applying~\lem{ito_formula} to~\eqn{sde_lyap}:
    \begin{subequations}\label{eq:levy_contract_eq_0}
        \begin{align}
            V(t,\qvect,\delta\qvect) &= V(s,\qvect,\delta\qvect) + \int_{s+}^t \partial_{\tau} V(\tau-,\qvect,\delta\qvect)d\tau\label{eq:levy_ito_det_part1}\\
            &\hskip1cm+ \int_{s+}^t\sum\limits_{i=1}^n \left[ \partial_{q_i}V(\tau-,\qvect,\delta\qvect) f_i(\tau,\qvect) + \partial_{\delta q_i}V(\tau-,\qvect,\delta\qvect) \left( F\delta\qvect\right)_i\right]d\tau \label{eq:levy_ito_det_part2} \\
            &\hskip1cm+ \int_{s+}^t\sum\limits_{i=1}^n \left[ \partial_{q_i}V(\tau-,\qvect,\delta\qvect) \sigma_{\mu,i}(\tau,\qvect) + \partial_{\delta q_i}V(\tau-,\qvect,\delta\qvect) \delta\sigma_{\mu,i}\right] dW(\tau)\label{eq:levy_ito_white_part}\\
            &\hskip1cm+ \frac{1}{2}\left[ \int_{s+}^t\sum\limits_{i,j=1}^n \partial_{\delta q_i\delta q_j}^2 V(\tau-,\qvect,\delta\qvect)d\left[ \delta q_i,\delta q_j\right]^c(\tau)\right.\label{eq:levy_ito_qv_part_1}\\
            &\hskip1cm\left. +\hskip.1cm 2\int_{s+}^t\sum\limits_{i,j=1}^n \partial_{q_i\delta q_j}^2 V(\tau-,\qvect,\delta\qvect)d\left[ q_i,\delta q_j\right]^c(\tau)\right.\label{eq:levy_ito_qv_part_2}\\
            &\hskip1cm\left. + \int_{s+}^t\sum\limits_{i,j=1}^n \partial_{q_iq_j}^2 V(\tau-,\qvect,\delta\qvect)d\left[ q_i,q_j\right]^c(\tau)\right]\label{eq:levy_ito_qv_part_3}\\
            &\hskip1cm + \sum\limits_{i=N(s)+1}^{N(t)} \left( V(T_i,\qvect,\delta\qvect) - V(T_i-,\qvect,\delta\qvect)\right)\label{eq:levy_ito_shot_part}
        \end{align}
    \end{subequations}
    where $F$ is the Jacobian from~\eqn{Fdef}, and $T_i$ denotes the time of the $i$th jump. As in the proof of~\thm{shot_contract}, a bound on~\eqn{levy_ito_det_part1} and~\eqn{levy_ito_det_part2} are derived from~\thm{basic_contract}.
    Simplifying the quadratic variation terms~\eqn{levy_ito_qv_part_1} to~\eqn{levy_ito_qv_part_3} requires computing the partial derivatives of $V$, which we omit the detailed calculations of for the sake of space.
    We obtain the following inequalities:
    \begin{subequations}\label{eq:levy_white_noise_part_bounds}
        \begin{align}
            &\sum\limits_{i,j = 1}^n \partial_{\delta q_i\delta q_j}^2 V(\tau-,\qvect,\delta\qvect) d[\delta q_i, \delta q_j]^c(\tau) \leq 2\overline{m}\gamma^2d\tau\label{eq:levy_white_noise_part_bounds_1}\\
            &\sum\limits_{i,j = 1}^n \partial_{q_i\delta q_j}^2 V(\tau-,\qvect,\delta\qvect) d[q_i, \delta q_j]^c(\tau) \leq m'\gamma^2\left( \int_0^1\norm{\partial_{\mu}\qvect(\mu,\tau)}^2 d\mu + 1\right)d\tau\label{eq:levy_white_noise_part_bounds_2}\\
            &\sum\limits_{i,j=1}^n \partial_{q_iq_j}^2 V(\tau-,\qvect,\delta\qvect) d[q_i, q_j]^c(\tau) \leq \left(m''\gamma^2\int_0^1\norm{\partial_{\mu}\qvect(\mu,\tau)}^2 d\mu\right)d\tau\label{eq:levy_white_noise_part_bounds_3}
        \end{align}
    \end{subequations}
    where $\gamma$ is the white-noise bound from~\defin{stoch_contract}, and $\overline{m}, m', m''$ are the metric bounds defined in~\assum{metric_bounds}. Note that we can use~\eqn{shot_contract_eq_7} to further simplify equations~\eqn{levy_ito_qv_part_1} to~\eqn{levy_ito_qv_part_3}. 
    Moreover, applying~\assum{theorem_assumptions} and following logic similar to that of the proof to~\thm{shot_contract} gives us a bound on~\eqn{levy_ito_shot_part}.
    
    Taking $\Ebb_{k}[\cdot]$ across the entire inequality, note that the white noise term~\eqn{levy_ito_white_part} disappears due to being a martingale with zero mean. 
    Combining the bounds of each remaining term from~\eqn{levy_contract_eq_0} yields the following:
    \begin{align}\label{eq:levy_contract_eq_5}
        \Ebb_{k}\left[ V(t,\qvect,\delta\qvect) \right] - \Ebb_k[V(s,\qvect,\delta\qvect)] &\leq -\left(2\alpha - \frac{m''\gamma^2}{2\underline{m}} - \frac{m'\gamma^2}{\underline{m}}\right) \int_{s+}^{t} \Ebb_{k}[V(\tau-,\qvect,\delta\qvect)] d\tau\notag\\
        &\hskip1cm + (m'\gamma^2 + \overline{m}\gamma^2)t + kh(t)
    \end{align}
    We obtain a bound on the solution $\Ebb_{k}[V(t,\qvect,\delta\qvect)]$ using~\lem{gronwall_negative_nonconstant} with $y(t) \triangleq \Ebb_k[V(t,\qvect,\delta\qvect)]$, $\zeta = kh(s)$, $\mu \triangleq 2\alpha - (m''\gamma^2/2\underline{m}) - (m'\gamma^2/\underline{m})$, and $\theta(t) \triangleq (m'\gamma^2 + \overline{m}\gamma^2)t + kh(t)$. Then we use~\eqn{shot_contract_eq_7} to write the resulting inequality in terms of the norm mean-squared error between $\xvect$ and $\yvect$.
    This gives us the desired bound~\eqn{levy_lyap_bound}.
\end{proof}

\begin{remark}
    The parameters~\eqn{levy_params} for the L\'{e}vy noise SDE~\eqn{combined_sde} can be expressed as a combination of~\eqn{white_params} and~\eqn{shot_params} in the following way. $\beta_{\ell}$ can be interpreted as being the direct sum of $\beta_w$ and $\beta_s (= 2\alpha)$ with the extra $2\alpha$ term removed to prevent double-counting of the convergence rate from the nominal system: $\beta_{\ell} = \beta_s + \beta_w - 2\alpha = \beta_w$.
    Furthermore, $\kappa_{\ell}(\beta_{\ell},s,t)$ is a sum of the white noise error ball and the shot noise error ball with contraction rate $\beta_{\ell}$ used in place of $\beta_w$ or $\beta_s$; this is written in the last two equations of~\eqn{kappa_levy}. 
    We emphasize the importance of this remark because of its likeness to the L\'{e}vy-Khintchine theorem, which represents L\'{e}vy processes as an additive combination of Brownian motion processes and compound Poisson processes.
\end{remark}

\begin{remark}
    The stochastic contraction theorems we presented in this section are comparable to the theories of hybrid systems or jump-Markov systems.
    In shot or L\'{e}vy noise systems, large deviations away from nominal behaviors arise solely from the jump-discontinuous noise process, which is independent of the open-loop dynamics. In contrast, hybrid systems have switches (i.e., jumps) which arise as an inherent property of the open-loop dynamics. Despite this important distinction, the two settings can still be closely related to one another in two ways. First, stability analysis techniques are primarily focused on handling the jump-discontinuities more than any other property of the system. For hybrid systems, literature towards this direction of research include Lyapunov-sense conditions for asymptotic stability~\cite{liberzon99lie,hespanha04} and characterizations of incremental stability~\cite{pham09hybrid}.
    Second, \textit{dwell time} can be related to the interarrival time by viewing it as a form of stability criteria which ensures that the system has sufficient time to converge towards a desired state in between consecutive switching phases. Likewise, the stability results of~\sec{stoch_contract} can be alternatively interpreted as conditions imposed on the shot or L\'{e}vy noise system such that the mean time between consecutive jumps (which depends on the intensity parameter $\lambda$) is long enough for the system to be reasonably close to the nominal trajectory.
    One notable example which utilizes dwell-time criteria for nonlinear systems is in Hespanha 1999~\cite{hespanha99}, where it is shown that input-to-state induced norms should be bounded uniformly between switches. In terms of applications, dwell-time criteria for attaining exponential stability has been shown to be effective for robotic systems, in particular walking locomotion and flapping flight~\cite{dorothy16} as well as autonomous vehicle steering~\cite{mesquita12}.
\end{remark}

\subsection{For Linear Time-Varying Systems}\label{subsec:shot_contract_linear}
To demonstrate a concrete example of the function $h(t)$ from~\eqn{time_var_bound}, we consider a specialization of~\thm{shot_contract} to the class of linear time-varying (LTV) systems where $f(t,\xvect) = A(t)\xvect(t)$ is a linear function and $\xi(t,\xvect) \equiv \xi(t)$ is a function is only dependent on time:
\begin{align}\label{eq:shot_sde_linear}
    d\xvect(t) = A(t)\xvect(t)dt + \xi(t)dN(t)
\end{align}
Here, $A:\Rbb^{+}\to\in\Rbb^{n\times n}$ is continuous for all $t\geq 0$, and $\xi:\Rbb^{+}\to\Rbb^n$ is a random function which maps time to a random vector in $\Rbb^n$ such that the bound in~\assum{bdd_noise} is still satisfied. By virtue of~\remk{abstract_c}, we can leverage the additional knowledge that the shot noise system is LTV in order to further simplify the bound~\eqn{cz0_exist}.

Note that a solution trajectory of~\eqn{shot_sde_linear} with value $\xvect(s)\in\Rbb^n$ at time $s \leq t$ can be explicitly written as:
\begin{align}\label{eq:shot_sde_linear_soln}
    \xvect(t) &= \Phi(t,s)\xvect(s) + \int_{s+}^t \Phi(t,\tau)\xi(\tau)dN(\tau) = \Phi(t,s)\xvect(s) + \sum\limits_{i=N(s)+1}^{N(t)} \Phi(t,T_i)\xi(T_i)
\end{align}
where the second equality follows from the definition of the Poisson integral from Section 2.3.2 of Applebaum 2009~\cite{applebaum09_book}, $N(t)$ is the number of jumps observed by time $t$, and $T_i \leq t$ denotes the time of the $i$th jump.

Instead of using a parameter $\mu\in[0,1]$, we construct the virtual system by stacking the SDEs~\eqn{shot_sde_linear} and the nominal system $d\yvect(t) = A(t)\yvect(t)dt$.
\begin{align}\label{eq:virtual_system_linear}
    d\qvect(t) = \begin{bmatrix} d\xvect(t)\\ d\yvect(t)\end{bmatrix} = \begin{bmatrix} A(t)\xvect(t)\\ A(t)\yvect(t)\end{bmatrix}dt + \begin{bmatrix} \xi(t)\\ 0 \end{bmatrix}dN(t)
\end{align}
where $N(t)$ is the standard Poisson process with intensity $\lambda>0$.

\begin{assumption}\label{assum:riccati_tv}
    There exists a continuously-differentiable, uniformly positive-definite, symmetric matrix $P(t)$ such that 
    \begin{enumerate}
        \item We can define the bounds
        \begin{align}
            \alpha_1 \triangleq \inf_t \lambda_{\text{min}}P(t), \quad \alpha_2 \triangleq \sup_t \lambda_{\text{max}}P(t)
        \end{align}
        for finite constants $\alpha_1,\alpha_2 > 0$.
        
        \item For all $t\geq 0$ and a fixed $\alpha > 0$,
        \begin{align}\label{eq:riccati_tv}
            \partial_tP(t) + P(t)A(t) + A^T(t)P(t) \leq -2\alpha P(t)
        \end{align}
    \end{enumerate}
\end{assumption}

The decomposition of $f(t,\xvect)$ into $A(t)\xvect$, which can be seen as a product of a function of time and a function of state, allows us to consider metrics $M(t,\qvect) \equiv M(t)$ which are independent of state $\qvect(t)\in\Rbb^{2n}$ of the virtual system~\eqn{virtual_system_linear}. 
Moreover, we choose the metric to be $M(t) \triangleq P(t)$, where $P(t)$ satisfies~\assum{riccati_tv}. 
The condition~\eqn{riccati_tv} can be viewed as a simplification of~\eqn{norm_squared_cond} for LTV systems. 
Because the metric is independent of state, we can construct a Lyapunov-like function which is simplified compared to~\eqn{sde_lyap}:
\begin{align}\label{eq:sde_lyap_linear}
    V(t,\qvect(t)) = (\yvect(t) - \xvect(t))^TP(t)(\yvect(t) - \xvect(t))
\end{align}

We further assume that the nominal system admits a solution with the state transition matrix $\Phi(t,\tau)\triangleq e^{\int_{\tau}^t A(r)dr}$ satisfying the following.

\begin{assumption}[Bounded State-Transition Matrix]\label{assum:shot_sdc_state_transition_mtx}
    The state-transition matrix $\Phi(t,\tau) \triangleq e^{\int_{\tau}^t A(r)dr}$ from any system with nominal dynamics $d\qvect(t) = A(t)\qvect(t)dt$ satisfies the following condition
    \begin{align}\label{eq:state_trans_bound}
        \norm{\Phi(t,\tau)}\leq\kappa e^{-\beta(t-\tau)},\qquad \forall \ 0\leq \tau\leq t
    \end{align}
    for some $\kappa,\beta > 0$.
\end{assumption}

\begin{theorem}[Shot Noise Stochastic Contraction Theorem: LTV Systems]\label{thm:shot_contract_linear}
    Suppose the LTV shot noise system~\eqn{shot_sde_linear} is perturbed by noise processes which satisfy~\assum{bdd_noise}, and is stochastically contracting in the sense of~\defin{stoch_contract} under the metric $P(t)$ from~\assum{riccati_tv}.
    Further suppose that the nominal LTV system is such that~\assum{shot_sdc_state_transition_mtx} and~\assum{riccati_tv} holds. 
    If for a fixed interval of time $[s,t]$ for $0\leq s<t$, $k\in\Nbb$ jumps occur with probability $p_k(t-s)$ given by~\eqn{poisson_prob}, then~\eqn{supermartingale_ineq_error} can be written explicitly as:
    \begin{align}\label{eq:shot_linear_lyap_bound}
        \Ebb_k[\norm{\yvect(t) - \xvect(t)}^2] \leq \frac{1}{\alpha_1}\Ebb_k[\norm{\yvect(s) - \xvect(s)}^2]e^{-\beta_s(t-s)} + \frac{\kappa_s(\beta_s,s,t)}{\alpha_1}
    \end{align}
    where \begin{subequations}\label{eq:shot_linear_params}
        \begin{align}
            \beta_s &\triangleq 2\alpha\label{eq:beta_shot_linear}\\
            \kappa_s(\beta_s,s,t) &\triangleq \int_{s+}^t\frac{d\psi_k(s,\tau)}{d\tau} e^{-\beta_s(t-\tau)}d\tau + k\alpha_2\eta^2e^{-\beta_s(t-s)}
            \label{eq:kappa_shot_linear}
        \end{align}
    \end{subequations}
    and
    \begin{align}\label{eq:psi_ebb}
        \psi_k(s,t) \triangleq \Ebb_k\left[\sum\limits_{i=1}^{k}\left\{2\alpha_2\eta\kappa \norm{\yvect(s) - \xvect(s)}e^{-\beta (T_i-s)} + 2\alpha_2\kappa\eta^2 \left(\sum\limits_{j=1}^{i-1} e^{-\beta (T_i-T_j)}\right)\right\}\right]
    \end{align}
    Here, $N(t)$ is the standard Poisson process in~\eqn{shot_sde}, and $T_i \geq s$, $i\geq 1$ is the arrival time of the $i$th jump after time $s$ in the Poisson process $N$ driving~\eqn{shot_sde_linear}, with convention $T_0 \triangleq s$. The variable $\lambda$ is the intensity of $N(t)$, $\alpha_1$ and $\alpha_2$ are defined in~\assum{riccati_tv}, and $\eta$ is the bound on the norm of the jumps $\xi(t)$ described by~\assum{bdd_noise}.
\end{theorem}

\begin{proof}[Proof of~\thm{shot_contract_linear}]
    Applying~\lem{ito_formula} to the Lyapunov-like function~\eqn{sde_lyap_linear} yields
\begin{subequations}\label{eq:shot_contract_linear_eq_0}
    \begin{align}
        V(t,\qvect) - V(s,\qvect) &= \int_{s+}^t\partial_{\tau}V(\tau-,\qvect)d\tau + \int_{s+}^t \left( \nabla_{\xvect}V(\tau-,\qvect)\cdot A(\tau)\xvect(\tau) + \nabla_{\yvect}V(\tau-,\qvect)\cdot A(\tau)\yvect(\tau)\right) d\tau\label{eq:shot_contract_linear_eq_0_det}\\
        &\hskip1cm + \sum\limits_{i=N(s)+1}^{N(t)} \left( V(T_i,\qvect) - V(T_i-,\qvect)\right)\label{eq:shot_contract_linear_eq_0_shot}
    \end{align}
\end{subequations}
As in the proof to~\thm{shot_contract}, we again abuse the notation of the subscript $i$ in $T_i$ for both sums which range over $i = N(s) + 1$ to $N(t)$ and sums which range over $i=1$ to $k$ for the sake of simplicity.
We use the left-limit notation of Section~\ref{subsec:notation}, and
\begin{align}\label{eq:shot_contract_linear_eq_1}
    \nabla_{\xvect}V(\tau,\qvect) = -2(\yvect(\tau) - \xvect(\tau))^TP(\tau), \qquad \nabla_{\yvect}V(\tau,\qvect) = 2(\yvect(\tau) - \xvect(\tau))^TP(\tau)
\end{align}

Let $\xvect(t)$ be the solution trajectory described in~\eqn{shot_sde_linear_soln} with value $\xvect(s)\in\Rbb^n$ at time $s$. 
Further denote $\yvect(t) = \Phi(t,s)\yvect(s)$ to be the solution trajectory of the nominal system with value $\yvect(s)\in\Rbb^n$ at time $s$.
We can simplify each term in the sum~\eqn{shot_contract_linear_eq_0_shot} as follows:
\begin{align}\label{eq:shot_contract_linear_eq_2}
    &V(T_i,\qvect) - V(T_i-,\qvect) = -2\yvect^TP\xi(T_i) + (\xvect^TP\xvect(T_i) - \xvect^TP\xvect(T_i-))\notag\\
    &\hskip2cm = -2\yvect^TP\xi(T_i) + 2\xi^T(T_i)P(T_i)\Phi(T_i,s)\xvect(s) + 2\left(\sum\limits_{j=N(s)+1}^{i-1} \xi^T(T_i)\Phi^T(T_i,T_j)\right)P(T_i)\xi(T_i) + \xi^TP\xi(T_i)
\end{align}
where $\yvect^TP\xi(\tau)$ is shorthand notation for $\yvect^T(\tau)P(\tau)\xi(\tau)$ for any $\tau$, and likewise for other similar notation. 
The first equality comes from the fact that $\yvect(T_i) = \yvect(T_i-)$ for all $T_i\leq t$ due to its continuity.
The second equality is obtained by by virtue of $N(T_i-) = N(T_i)-1$ and $N(T_i) = i$.

Substituting~\eqn{shot_contract_linear_eq_2} into~\eqn{shot_contract_linear_eq_0} yields:
\begin{subequations}\label{eq:shot_contract_linear_eq_4}
    \begin{align}
        &V(t,\qvect) - V(s,\qvect) = \int_{s+}^t (\yvect(\tau) - \xvect(\tau-))^T(\partial_{\tau}P(\tau) + 2P(\tau)A(\tau))(\yvect(\tau) - \xvect(\tau-))d\tau\notag\\
        &\hskip1cm + \sum\limits_{i=N(s)+1}^{N(t)} \left[-2\xi^T(T_i)P(T_i)\Phi(T_i,s)(\yvect(s) - \xvect(s)) + 2\left(\sum\limits_{j=N(s)+1}^{i-1} \xi^T(T_i)\Phi^T(T_i,T_j)\right)P(T_i)\xi(T_i) + \xi^TP\xi(T_i)\right]\label{eq:shot_contract_linear_eq_4_1}\\
        &\leq -2\alpha\int_{s+}^t V(\tau-,\qvect)d\tau + \sum\limits_{i=N(s)+1}^{N(t)} \left[2\alpha_2\norm{\xi(T_i)}\norm{\Phi(T_i,s)}\norm{\yvect(s) - \xvect(s)} + 2\alpha_2\left(\sum\limits_{j=N(s)+1}^{i-1} \xi^T(T_j)\Phi^T(T_i,T_j)\right)\xi(T_i) + \alpha_2\xi^T\xi(T_i)\right]\label{eq:shot_contract_linear_eq_4_2}\\
        &\leq -2\alpha\int_{s+}^t V(\tau-,\qvect)d\tau + \sum\limits_{i=N(s)+1}^{N(t)}\bigg[2\alpha_2\eta\kappa \norm{\yvect(s) - \xvect(s)}e^{-\beta (T_i-s)} + 2\alpha_2\kappa\eta^2 \left(\sum\limits_{j=N(s)+1}^{i-1} e^{-\beta (T_i-T_j)}\right) + \alpha_2\eta^2\bigg]\label{eq:shot_contract_linear_eq_4_4}
    \end{align}
\end{subequations}
Here,~\eqn{shot_contract_linear_eq_4_2} from using~\assum{riccati_tv}, the bound on $P(t)$, and submultiplicativity.
The last inequality~\eqn{shot_contract_linear_eq_4_4} comes from triangle inequality,~\eqn{state_trans_bound}, and~\assum{bdd_noise}. 
Taking the conditional expectation $\Ebb_k[\cdot]$ over~\eqn{shot_contract_linear_eq_4} yields:
\begin{align}\label{eq:shot_contract_linear_eq_5}
    \Ebb_k[V(t,\qvect)] - \Ebb_k[V(s,\qvect)] &\leq -2\alpha\int_{s+}^t\Ebb_k\left[V(\tau-,\qvect)\right]d\tau + \psi_k(s,t) + k\alpha_2\eta^2
\end{align}
where $\psi_k(s,t)$ is as in~\eqn{psi_ebb}. Apply~\lem{gronwall_negative_nonconstant} to~\eqn{shot_contract_linear_eq_5} with $y(t) \triangleq \Ebb_k[V(t,\qvect)]$, $\zeta \triangleq k\alpha_2\eta^2$, $\mu \triangleq 2\alpha$, and $\theta(t) \triangleq \psi_k(s,t)$. 
Use~\eqn{shot_contract_eq_7} and the fact that $(1/\alpha_2)I_n \leq P^{-1}(t) \leq (1/\alpha_1)I_n$ to write the inequality in terms of the norm mean-squared error between $\xvect$ and $\yvect$.
This gives us the desired bound~\eqn{shot_linear_lyap_bound}, with $\beta_s$ as in~\eqn{beta_shot_linear} and $\kappa_s(\beta_s,t)$ as in~\eqn{kappa_shot_linear}.
\end{proof}

There are several ways to simplify $\psi_k(s,t)$.
One can explicitly write out the integral form of the expectation with the knowledge that $T_i$ are Gamma-distributed with parameter $i$ and $\lambda$ for all $i = 1, \cdots, k$.
While this direct computation of~\eqn{psi_ebb} yields the tightest bound, this method requires computing $k$ integrals and is thus increasingly difficult to compute with increasing $k$.
For concreteness, we look at two specific ways to derive a looser bound.

The first term of~\eqn{psi_ebb} simplifies as
\begin{align}\label{eq:shot_contract_linear_eq_5_5}
    2\alpha_2\eta\kappa\Ebb_k\left[\sum\limits_{i=1}^k \norm{\yvect(s) - \xvect(s)}e^{-\beta(T_i - s)}\right] &\leq 
    2\alpha_2\eta\kappa e^{\beta s}\Ebb_k[\norm{\yvect(s) - \xvect(s)}]k\Ebb_k\left[\max_{i=1,\cdots, k} e^{-\beta T_i}\right] \notag\\
    &= 2\alpha_2\eta\kappa e^{\beta s}\Ebb_k[\norm{\yvect(s) - \xvect(s)}]k\Ebb\left[e^{-\beta T_1}\right] 
\end{align}
because the random variables $T_i, i=1,\cdots,k$ are such that $T_1\leq T_2\leq\cdots T_k$, and so $e^{-\beta T_i}$ takes the largest value with the smallest index $i$. 

For the second term of~\eqn{psi_ebb}, we can invoke~\lem{max_nng}:
\begin{align}\label{eq:second_inequality_max_max}
    2\alpha_2\kappa\eta^2\Ebb_k\left[\sum\limits_{i=1}^k\sum\limits_{j=1}^{i-1} e^{-\beta(T_i - T_j)}\right] &\leq 2\alpha_2\kappa\eta^2k(k-1)\Ebb_k\left[\max_{i=1,\cdots, k} e^{-\beta T_i} \max_{j=1,\cdots, k} e^{\beta T_j}\right] = 2\alpha_2\kappa\eta^2k(k-1)\Ebb_k\left[e^{-\beta T_1} e^{\beta T_k}\right]
\end{align}
More specifically,~\eqn{second_inequality_max_max} holds because $T_i, i=1,\cdots,k$ are such that $T_1\leq T_2\leq\cdots T_k$. 
This means the value of $i\in\{1,\cdots, k\}$ which maximizes $e^{-\beta T_i}$ is $i=1$, and the value of $j\in\{1,\cdots, k\}$ which maximizes $e^{\beta T_j}$ is $j=k$.
Use the fact that $T_1\sim\text{Exp}(\lambda)$ and $T_k\sim\text{Gamma}(k,\lambda)$ to further simplify the resulting inequality.
The derivative of the function $\psi_k(s,t)$ from~\eqn{psi_ebb} then simplifies to
\begin{align}\label{eq:max_max_psi}
    \frac{d\psi_k(s,t)}{dt} \triangleq 2\alpha_2\eta\kappa \norm{\yvect(s) - \xvect(s)}ke^{\beta s}\lambda e^{-(\lambda+\beta)t} + \frac{2\alpha_2\kappa\eta^2 k(k-1)\lambda^{k+1}}{(\lambda+\beta)\Gamma(k)}t^{k-1}\left(e^{(\beta - \lambda) t} - e^{-2\lambda t}\right)
\end{align}

Alternatively, we can bound the second term of~\eqn{psi_ebb} in the following way:
\begin{align}\label{eq:second_inequality_sum_exp}
    2\alpha_2\kappa\eta^2\Ebb_k\left[\sum\limits_{i=1}^k\sum\limits_{j=1}^{i-1} e^{-\beta(T_i - T_j)}\right] &\leq 2\alpha_2\kappa\eta^2\Ebb_k\left[\sum\limits_{i=1}^k(i-1)\max_{j\in\{1,\cdots,i-1\}}e^{-\beta (T_i-T_j)}\right] = \alpha_2\kappa\eta^2k(k-1)\Ebb_k\left[e^{-\beta S_i}\right]
\end{align}
where $S_i \triangleq T_i - T_{i-1}$ is exponentially-distributed with parameter $\lambda$.
The maximum value of $j$ is achieved at $j = i-1$ because the difference $T_i-T_{i-1}$ is the smallest value in the range, and thus $e^{-\beta(T_i-T_{i-1})}$ is the largest.
Note that $S_i$ represent the interarrival times of the Poisson jumps and are thus i.i.d. for all $i = 1, \cdots, k$.
The derivative of the function $\psi_k(s,t)$ from~\eqn{psi_ebb} then simplifies to
\begin{align}\label{eq:sum_exp_psi}
    \frac{d\psi_k(s,t)}{dt} \triangleq \left(2\alpha_2\eta\kappa \norm{\yvect(s) - \xvect(s)}ke^{\beta s} + \alpha_2\kappa\eta^2 k(k-1)\right)\lambda e^{-(\lambda+\beta)t}
\end{align}

As mentioned in~\remk{tradeoff}, we see that $\psi_k(s,t)$ of~\eqn{psi_ebb} (and thus $\kappa_s(\beta_s,s,t)$ of~\eqn{kappa_shot_linear}) is directly proportional to the jump norm bound $\eta$.

\begin{remark}\label{rmk:shot_ltv_compare}
    A comparison of the results between~\thm{shot_contract_linear} and~\thm{shot_contract} show that the form of the stability bounds are the same. First, note that $\underline{m} = \alpha_1$ and $\beta_s$ from~\eqn{beta_shot_linear} in~\thm{shot_contract_linear} is as in~\eqn{beta_shot} from~\thm{shot_contract}.
    Second, and more importantly, having more knowledge about the system dynamics allows us to derive a more concrete bound compared to the bounds of Section~\ref{subsec:stoch_contract_thms}, which are dependent upon some abstract function $h(t)$.
    In particular, for this LTV case, the difference~\eqn{cz0_exist} can be computed exactly using the precise solution form~\eqn{shot_sde_linear_soln}, and the metric $M(t,\xvect)\triangleq P(t)$ does not depend on the state $\xvect$.
    From~\eqn{shot_contract_linear_eq_5}, we have $h(t) = \psi_k(s,t)$.
\end{remark}

\begin{remark}\label{rmk:tradeoff_linear}
    Similar to~\remk{tradeoff}, the strength of the stability bound in~\thm{shot_contract_linear} is contingent on both inherent parameters and design parameters, but the explicit form of~\eqn{psi_ebb} allows us to derive additional insights.
    First, an additional inherent parameter we can consider is the the maximum norm bound $\eta$:~\eqn{psi_ebb} is directly proportional to $\eta$, which implies that larger jump norms result in larger error bounds.
    Second, among the design parameters, the contraction metric $M(t,\xvect)\equiv P(t)$ can be chosen such that $\alpha_1$ is small and $\alpha_2$ is large.
    This enables a looser bound on~\eqn{shot_linear_lyap_bound}, and so~\eqn{shot_linear_lyap_bound} is tighter and more meaningful when we choose a metric whose condition number is close to $1$.
    For exponentially-stable unperturbed LTV systems,~\eqn{psi_ebb} also demonstrates that a larger deterministic contraction rate allows for faster convergence to a smaller error ball.
    Correspondingly, this effect can also be achieved for controllable open-loop unstable unperturbed LTV systems by designing a control law such that, in~\assum{shot_sdc_state_transition_mtx}, $\kappa$ is small and $\beta$ is large.
    In~\sec{simulation}, we use numerical simulations to investigate how the stability bounds vary by varying the different parameters discussed in here and in~\remk{tradeoff}.
\end{remark}


\section{Numerical Examples}\label{sec:simulation}
\subsection{2D Nonlinear System}\label{subsec:2D_nonlinear}
We demonstrate the general bound from the Shot Noise Stochastic Contraction~\thm{shot_contract} while considering the following 2D nonlinear shot noise system:
\begin{align}\label{eq:2D_nonlinear}
    d\xvect(t) = d\begin{bmatrix}x_1\\ x_2\end{bmatrix}(t) 
    = \underbrace{\begin{bmatrix}f_1(t,\xvect)\\ f_2(t,\xvect)\end{bmatrix}}_{=: f(t,\xvect)}dt + \underbrace{\begin{bmatrix}\xi_1(t)\\ \xi_2(t)\end{bmatrix}}_{=: \xi(t)}dN(t)
\end{align}
with $\sup_{t>0}\norm{\xi(t)}\leq \eta$, for some $\eta > 0$.
We choose a system whose nominal dynamics are open-loop exponentially stable, because the problem of stabilizing controller design for unperturbed systems is not the focus of this section.
We also choose a nonlinear function $f(t,\xvect)$ such that its Jacobian matrix $F$ turns into a diagonal matrix, and such that~\eqn{norm_squared_cond} is satisfied with a state-dependent $M(t,\xvect)$. One such dynamics are given by
\begin{align}
    f(t,\xvect) \triangleq \begin{bmatrix}-(2t+1)(\sin(x_1) + 3)x_1\\ -(2t+1)(\cos(x_2) + 5)x_2\end{bmatrix}
\end{align}
which has diagonal matrix Jacobian
\begin{align}
    F(t,\xvect) \triangleq \nabla_{\xvect}f(t,\xvect) = \begin{bmatrix}
        -(2t+1)\left[x_1\cos(x_1) + \sin(x_1) + 3\right] & 0\\ 0 & -(2t+1)\left[-x_2\sin(x_2) + \cos x_2 + 5\right]
    \end{bmatrix}
\end{align}

Choose the metric
\begin{align}
    M(t,\xvect) &\triangleq \begin{bmatrix}
        (\sin(t) + 3)(e^{-x_1^2} + 1) & 0\\ 0 & (\cos(t) + 3)(e^{-x_2^2} + 1)
    \end{bmatrix}
\end{align}
We numerically verify that the conditions of~\thm{basic_contract} hold, and so the nominal system is incrementally stable under the metric $M(t,\xvect)$ defined above.
Moreover, $M(t,\xvect)$ is uniformly positive definite with $\overline{m} = 4, \underline{m} = 2$.

We empirically generate trajectories with initial condition $\xvect_0 \sim U[1,6]$ up until the maximum time $t$ where the number of jumps is $N(t) = 3$ such that~\eqn{norm_squared_cond} is satisfied.
To construct the virtual system, we choose the specific affine parametrization $\qvect(\mu,t) = (1-\mu)\xvect(t) - \mu\yvect(t), \partial_{\mu}\qvect(t) = \yvect(t) - \xvect(t)$. Then the Lyapunov-like function~\eqn{sde_lyap} is written explicitly as:
\begin{align}\label{eq:2D_nonlinear_sde_lyap}
    &V(t, \qvect, \delta\qvect) = \int_0^1 \partial_{\mu}\qvect^TM(t,\qvect(\mu,t))\partial_{\mu}\qvect d\mu\notag\\
    &\hskip.5cm = \int_0^1 (\yvect(t) - \xvect(t))^T M\left(t,(1-\mu)\xvect(t) - \mu\yvect(t)\right)(\yvect(t) - \xvect(t)) d\mu\notag\\
    &\hskip.5cm = \int_0^1 \left[(y_1(t) - x_1(t))^2(\sin(t) + 3)\left(e^{-((1-\mu)x_1(t) - \mu y_1(t))^2} + 1\right) + (y_2(t) - x_2(t))^2(\cos(t) + 3)\left(e^{-((1-\mu)x_2(t) - \mu y_2(t))^2} + 1\right)\right]d\mu
\end{align}

\begin{figure}
    \begin{center}
    \includegraphics[width=0.7\columnwidth]{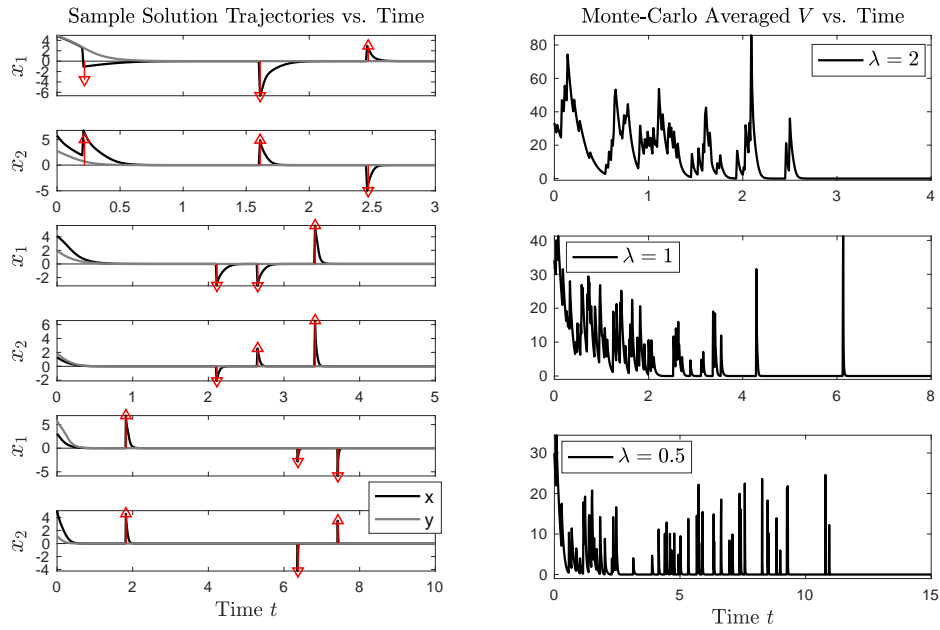}
    \caption{
    Simulation of the exponentially stable 2D nonlinear system~\eqn{2D_nonlinear} for varying values of $\lambda\in\{0.5,1,2\}$. [Left column] Evolution of a sample solution trajectories $\xvect$ and $\yvect$ with the corresponding value of $\lambda$ over time. [Right column] Lyapunov-like function $V$ averaged over $20$ Monte-Carlo trials, then plotted over time. The number of jumps is fixed at $N(t) = 3$.
    }
    \label{fig:2D_nonlinear_stable}
    \end{center}
\end{figure}

With the Lyapunov-like function~\eqn{2D_nonlinear_sde_lyap}, each term in the sum of~\eqn{shot_contract_eq_0_5_shot} yields:
\begin{align}\label{eq:2D_nonlinear_sde_lyap_diff}
    &V(T_i, \qvect(T_i), \delta\qvect(T_i)) - V(T_i-, \qvect(T_i-), \delta\qvect(T_i-))\notag\\
    &= \int_0^1 (\sin(T_i) + 3)\left[(y_1(T_i) - x_1(T_i))^2\left(e^{-((1-\mu)x_1(T_i) - \mu y_1(T_i))^2} + 1\right) - (y_1(T_i) - x_1(T_i-))^2\left(e^{-((1-\mu)x_1(T_i-) - \mu y_1(T_i))^2} + 1\right)\right]d\mu\notag\\
    &\hskip.5cm + \int_0^1 (\cos(T_i) + 3)\left[(y_2(T_i) - x_2(T_i))^2\left(e^{-((1-\mu)x_2(T_i) - \mu y_2(T_i))^2} + 1\right) - (y_2(T_i) - x_2(T_i-))^2\left(e^{-((1-\mu)x_2(T_i-) - \mu y_2(T_i))^2} + 1\right)\right]d\mu
\end{align}
for each arrival time $T_i>0$.
The evolution of some sample solution trajectories and an empirical average over multiple Monte-Carlo trials for three different values of $\lambda$ are portrayed in~\fig{2D_nonlinear_stable}.
Because we cannot compute the explicit form of $\xvect(t)$ and $\yvect(t)$ from the dynamics, the bound $h(t)$ from~\eqn{time_var_bound} is difficult to compute analytically. 
However, we can observe empirically from~\fig{2D_nonlinear_stable} that as $\lambda$ decreases, the support of $V$ increases over time, i.e., the spikes spread out over longer intervals of time with smaller $\lambda$ values.
Additionally, the mean height of the spikes decreases, which experimentally verifies~\remk{tradeoff}.

\subsection{1D Linear Reference-Tracking}\label{subsec:1D_lqr}
We now use the Shot Noise Stochastic Contraction Theorem to derive a stability bound for a simple linear system perturbed by shot noise, as a further specialization to the LTV system from Section~\ref{subsec:shot_contract_linear}. 
Suppose we have the following scalar system, which can be viewed as the Ornstein-Uhlenbeck process augmented with shot noise instead of white noise:
\begin{align}\label{eq:1D_system}
    dx(t) = ax(t)dt + u(t)dt + \xi(t)dN(t)
\end{align}
where $a > 0$ so that the system is unstable in open-loop, $N(t)$ is a standard Poisson process with rate $\lambda > 0$, and jump height distribution $\xi(t)$ is a Bernoulli random variable which takes value $\eta > 0$ with probability $p$, and $-\eta$ with probability $q \triangleq 1 - p$. 

We are interested in the problem of tracking some given reference trajectory $x_r(t)$. 
To achieve this, we design the following control law:
\begin{align}\label{eq:ref_track_ctrl}
    u_{r}(t) = \dot{x}_r(t) - ax_r(t), \quad u(t) = u_{r}(t) - \gamma(x(t) - x_r(t))
\end{align}
with control gain $\gamma > a$.
As described in Section~\ref{subsec:notation}, the dot notation of $\dot{x}_r(t)$ refers to the time-derivative of $x_r(t)$.
The system~\eqn{1D_system} can be solved directly:
\begin{align}\label{eq:1D_system_soln}
    x(t) &= x_0e^{(a-\gamma)t} + \int_0^t \left(u_r(s) + \gamma x_r(s)\right)e^{(a-\gamma)(t-s)}ds + \sum\limits_{i=1}^{N(t)} \xi(T_i)e^{(a-\gamma)(t - T_i)}
\end{align}
where $x_0\in\Rbb$ is the initial condition. The nominal closed-loop system has the dynamics $dy(t) = ay(t) + u(t)$, with the same $u(t)$ as in~\eqn{ref_track_ctrl}. A trajectory of this nominal system with initial condition $y_0\in\Rbb$ is thus given by
\begin{align}\label{eq:1D_system_nom_soln}
    y(t) &= y_0e^{(a-\gamma)t} + \int_0^t \left(u_r(s) + \gamma x_r(s)\right)e^{(a-\gamma)(t-s)}ds
\end{align}

\begin{figure}
    \begin{center}
    \includegraphics[width=0.75\columnwidth]{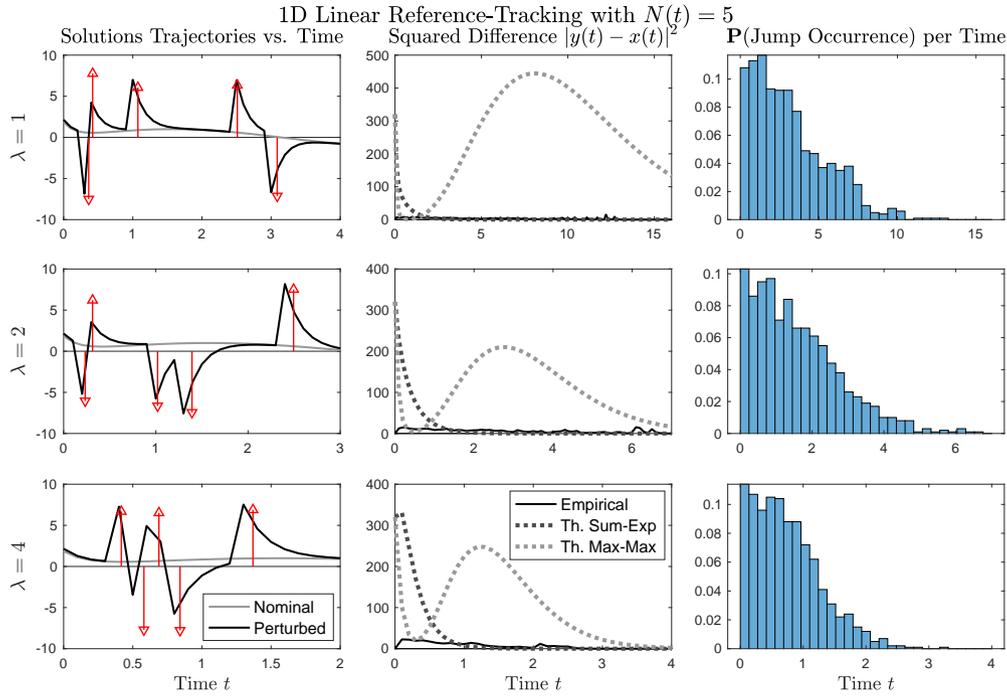}
    \caption{Simulation of the 1D linear reference-tracking system over values of $\lambda \in \{1,2,4\}$, constrained to $k=5$ jumps.
    [Left column] A sample trajectory of the nominal system (light grey line) and a trajectory of the shot noise perturbed system (black line). Jump occurrences and heights are marked with red stems.
    [Middle column] The empirical mean squared-difference is obtained by taking the timewise average over all Monte-Carlo trials of trajectory squared-differences.
    Two theoretical bounds are computed, one using $d\psi_k/dt$ from~\eqn{sum_exp_psi} (dark grey dashed line), one using $d\psi_k/dt$ from~\eqn{max_max_psi} (light grey dashed line).
    [Right column] The empirical probability that a jump occurs at a certain time is plotted over time.
    The histogram is constructed by discretizing the span of time into subintervals and computing the proportion of jumps over all Monte-Carlo trials which fall into each subinterval.
    }
    \label{fig:1D_lqr_avgV_probs}
    \end{center}
\end{figure}

Like~\eqn{virtual_system_linear}, we design a virtual system by stacking the nominal closed-loop system on top of the noise-perturbed closed-loop system~\eqn{1D_system}, and define $\qvect = (x, y)^T \in \Rbb^2$ to be the virtual system state. 
The contraction metric $P(t)$ from~\assum{riccati_tv} is chosen to be the identity $I_2$, meaning both upper and lower bounds $\alpha_1 = \alpha_2 = 1$. 
The Lyapunov-like function is chosen to be $V(\qvect) = (y - x)^2$.

Following an argument similar to the proof of~\thm{shot_contract_linear} with $t_0 = 0$, we get:
\begin{align}\label{eq:1D_linear_eq0}
    \Ebb_k[V(\qvect(t))] - \Ebb_k[V(\qvect_0)] &\leq
    2(a-\gamma)\int_{0+}^t \Ebb_k[V(\qvect(s))] ds + \Ebb_k\left[\sum\limits_{i=1}^k (y-x)^2(T_i) - (y-x)^2(T_i-)\right]
\end{align}
conditioned on the number of jumps being $N(t) = k$ by time $t$.
Here, the last term can be simplified by using~\eqn{1D_system_soln} and~\eqn{1D_system_nom_soln}:
\begin{align}\label{eq:1D_linear_eq1}
    (y-x)^2(T_i) - (y-x)^2(T_i-) &= -2(y_0 - x_0)e^{(a-\gamma)T_i}\xi(T_i) + 2\xi(T_i)\sum\limits_{j=1}^{i-1} \xi(T_j)e^{(a-\gamma)(T_i-T_j)} + \xi^2(T_i)
\end{align}
Substituting~\eqn{1D_linear_eq1} into~\eqn{1D_linear_eq0} and using~\thm{shot_contract_linear} yields the following bound with probability $p_k(t)$ given by~\eqn{poisson_prob}:
\begin{align}\label{eq:1D_kappa_part}
    &\Ebb_k\left[\abs{y(t) - x(t)}^2\right] \leq \Ebb_k\left[\abs{y_0 - x_0}^2\right]e^{-\beta_st} + \kappa_s(\beta_s,0,t)
\end{align}
where the contraction rate is $\beta_s \triangleq 2(\gamma-a) > 0$ and the error bound $\kappa_s(\beta_s,0,t)$ comes from~\eqn{kappa_shot}.

We simulate~\eqn{1D_system} with the lower-level tracking controller~\eqn{ref_track_ctrl} implemented to track $x_r(t) \triangleq \sin(t)$.
We use the theoretical bound created for the each of the two different versions of $\psi_k$,~\eqn{max_max_psi} and~\eqn{sum_exp_psi}.
The results are organized in~\fig{1D_lqr_avgV_probs}.
The intensity of the shot noise process varies across $\lambda \in \{1,2,4\}$, corresponding to each row of subfigures. 
All three rows share the following common experiment setup: the number of jumps is fixed to be $k = 5$, and we simulate the evolution of trajectories starting from $t_0 = 0$ until just before the $(k+1)$th jump occurs at time $T_{k+1}$.
In the left column of subfigures, we simulate a sample nominal closed-loop reference trajectory $y(t)$ (grey solid line) with a sample closed-loop noise-perturbed reference trajectory $x(t)$ (black solid line). 
In the middle column of subfigures, the empirical average squared-difference $\abs{y(t) - x(t)}^2$ is computed by timewise-averaging over $200$ Monte-Carlo trial trajectories (black solid line). 
The two types of the theoretical mean-squared error bound~\eqn{1D_kappa_part} are also plotted, using $d\psi_k/dt$ from~\eqn{sum_exp_psi} (dark grey dashed line) and from~\eqn{max_max_psi} (light grey dashed line).
In the right column of subfigures, the empirical probability that a jump occurs at a certain time is plotted as a histogram over time.
The histogram is constructed by discretizing the maximum length of time into $30$ subintervals and computing the proportion of jumps (over all $200$ Monte-Carlo trials) which fall into each subinterval.
We note that it is possible for the empirical squared-difference to exceed the theoretical bound.
This is because the theoretical error bound is on expected behavior, and should not be treated as an almost-sure guarantee for all sample paths.
Moreover, we observe that in both figures, the trajectories converge towards each other in between consecutive jumps, which aligns with the incrementally stable nature of the nominal system.

Compared to the previous simulation of Section~\ref{subsec:2D_nonlinear}, we can make a few insightful observations based on the two figures.
The theoretical bound derived in Section~\ref{subsec:shot_contract_linear} yields an expression for $\psi_k(0,t)$ which is proportional to both $\lambda$ and $e^{-\lambda t}$.
As seen in the middle column of subfigures of~\fig{1D_lqr_avgV_probs}, this effect is demonstrated for both the theoretical bound computed using~\eqn{sum_exp_psi} and~\eqn{max_max_psi}.
First, note that an increasing value of $\lambda$ corresponds to a larger accumulation of jumps.
This corresponds to an increasing constant initial value in both grey dashed lines.
Second, an increasing value of $\lambda$ also corresponds to a faster accumulation of jumps, i.e. all $k=5$ jumps of the system occur earlier in time for larger $\lambda$.
This corresponds to a faster speed of decay in the first bumps of both grey dashed lines.
Moreover, for the light grey dashed line, the effect of $\lambda$ is illustrated through the proximity between the line $t = 0$ and the second bump; as $\lambda$ grows larger, more jumps occur earlier in time, and the second bump occurs closer to $t = 0$.
These empirical results demonstrate what was qualitatively observed by~\remk{tradeoff_linear}.
Another observation is that the second bump which occurs when computing the theoretical bound using~\eqn{max_max_psi} (light grey dashed line) arises from the distribution of $T_k$.
The second bump essentially accounts for the possibility of seeing jumps which occur closer to $t$ in the interval $[0,t]$.
In contrast, the theoretical error bound with $\psi_k(t)$ as in~\eqn{sum_exp_psi} only has the initial bump; all the weight is assigned to the initial value, from which it exponentially decays over time.

\subsection{2D LTV Systems}\label{subsec:2D_ltv}
We extend the experiment of Section~\ref{subsec:1D_lqr} by considering more complex 2D LTV shot noise systems of the form:
\begin{align}
    d\xvect(t) = d\begin{bmatrix}x_1\\ x_2\end{bmatrix}(t) 
    = \underbrace{\begin{bmatrix} a_{11}(t) & a_{12}(t)\\ a_{21}(t) & a_{22}(t)\end{bmatrix}}_{=: A(t)}\begin{bmatrix}x_1\\ x_2\end{bmatrix}(t)dt + \underbrace{\begin{bmatrix}\xi_1(t)\\ \xi_2(t)\end{bmatrix}}_{=: \xi(t)}dN(t)
\end{align}
with $\sup_{t>0}\norm{\xi(t)}\leq \eta$ for some $\eta > 0$. 
Note that constructing virtual system~\eqn{virtual_system_linear} for 2D dynamics yields virtual system state vector $\qvect(t) \in\Rbb^4$. 
For a more practical setup, we can follow the design of the previous 1D example from Section~\ref{subsec:1D_lqr} and consider an open-loop unstable system $A(t)$ with a control law of the form $u(t) \triangleq K(t)\xvect(t)$ such that the closed-loop system is exponentially stable. 
However, for the simplicity of the example, we do not consider the controller design problem, and demonstrate the contraction theorems on systems which are already open-loop exponentially stable. 

We apply Section~\ref{subsec:shot_contract_linear} to derive the theoretical mean-squared error bounds for 2D LTV systems with one of two types of matrices $A(t)$: diagonal and (upper) triangular.
We note that the relationship between increasing $\lambda$ and the variation in error bound for both types of systems is similar to what was observed in Section~\ref{subsec:1D_lqr} regardless of the two different approaches (\eqn{sum_exp_psi} versus~\eqn{max_max_psi}) to simplifying the error bound.
Hence, in contrast to the experiment of Section~\ref{subsec:1D_lqr}, we illustrate the results for only one choice of $\lambda$.
The primary purpose of the simulations in this section is to demonstrate the analytical computation of the bounds for more complex LTV systems, especially in choosing the metric $P(t)$ and all the appropriate parameters values to satisfy the assumptions of~\thm{shot_contract_linear}.

\paragraph{Diagonal $A(t)$ Matrix}
First, consider the case where $A(t)$ is a diagonal matrix, i.e., $a_{12}(t) = a_{21}(t) \equiv 0$ for all $t\geq 0$.
Specifically, choose:
\begin{align}
    A(t) &\triangleq \begin{bmatrix} -3t^2-1 & 0 \\ 0 & -2t-1 \end{bmatrix}
\end{align}

Because $A(t)$ is diagonal, the corresponding state-transition matrix is easily computed:
\begin{align}\label{eq:2D_diag_ltv_stable_bddP_Phi}
    \Phi(t,s) \triangleq \text{exp}\left(\begin{bmatrix}\int_s^t a_{11}(r)dr & 0\\ 0 & \int_s^t a_{22}(r)dr \end{bmatrix}\right) = \begin{bmatrix}e^{-(t^3 + t) + (s^3 + s)} & 0\\ 0 & e^{-(t^2 + t) + (s^2 + s)}\end{bmatrix}
\end{align}
One choice of parameters such that~\assum{shot_sdc_state_transition_mtx} is satisfied with the induced $2$-norm is when $\kappa = 1$ and $\beta = 1$.

Solution trajectories for the perturbed and nominal systems can now be written explicitly as follows. From~\eqn{shot_sde_linear_soln}, we have the perturbed system trajectory $\xvect(t)$ and the nominal trajectory $\yvect(t)$ given by:
\begin{align}\label{eq:2D_diag_ltv_stable_bddP_solns}
    \xvect(t) &= \begin{bmatrix} e^{-(t^3 + t)}x_{0,1}\\ e^{-(t^2 + t)}x_{0,2}\end{bmatrix} + \sum\limits_{i=1}^{N(t)} \begin{bmatrix} e^{-(t^3 + t) + (T_i^3 - T_i)}\xi_1(T_i)\\ e^{-(t^2 + t) + (T_i^2 - T_i)}\xi_2(T_i)\end{bmatrix}, \quad \yvect(t) = \begin{bmatrix} e^{-(t^3 + t)}y_{0,1}\\ e^{-(t^2 + t)}y_{0,2}\end{bmatrix}
\end{align}
with initial conditions $\xvect_0,\yvect_0\in\Rbb^2$, respectively.

\begin{figure}
    \begin{center}
    \includegraphics[width=0.6\columnwidth]{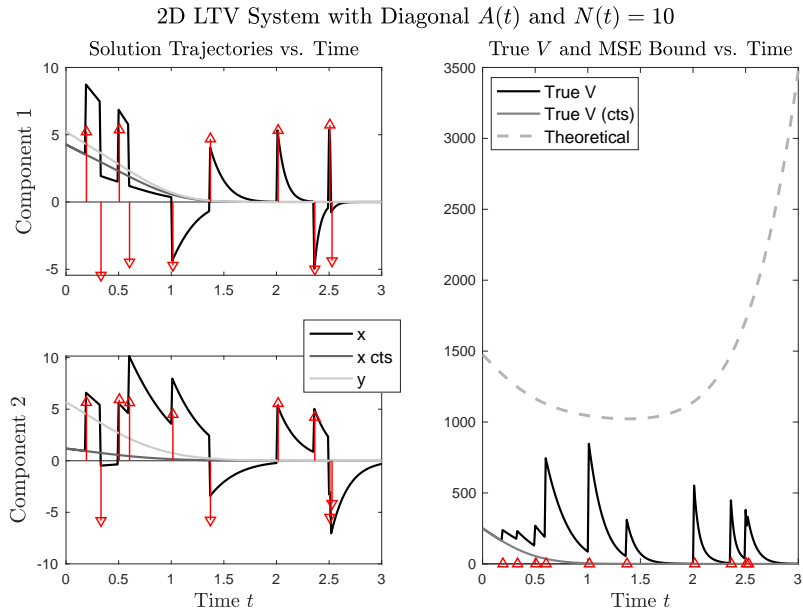}
    \caption{Simulation of the exponentially stable 2D diagonal LTV system with bounded metric $P(t)$. 
    The number of jumps is fixed at $N(t) = 10$, indicated using red stems.
    [Left column] Evolution of solution trajectories $\xvect(t)$ (black), $\xvect^c(t)$ (dark grey), and $\yvect$ (light grey) over time. 
    [Right] Evolution of the empirical Lyapunov-like function $V$ (black solid), $V^c$ (dark grey solid), and the theoretical bound, computed with $d\psi_k/dt$ as in~\eqn{max_max_psi}, (light grey dashed) over time. 
    }
    \label{fig:2D_diagonal_ltv_stable_bddP}
    \end{center}
\end{figure}

Note that one such choice of metric $P$ which satisfies~\eqn{riccati_tv} is:
\begin{align}
    P(t) &\triangleq \begin{bmatrix} \sin(t) + 3 & 0 \\ 0 & \cos(t) + 3\end{bmatrix}
\end{align}
with $\alpha = 2$. Clearly, $P(t)$ is positive definite for each $t > 0$, and satisfies the boundedness inequality in~\assum{riccati_tv} with $\alpha_1 = 2$ and $\alpha_2 = 4$. Hence, the Lyapunov-like function~\eqn{sde_lyap_linear} yields:
\begin{align}
    V(t,\qvect(t)) &= (\sin(t) + 3)(y_1(t) - x_1(t))^2 + (\cos(t) + 3)(y_2(t) - x_2(t))^2\notag\\
    &= V^c(t,\qvect(t)) + e^{-2(t^3 + t)}(\sin(t) + 3)\left\{ 2(y_{0,1} - x_{0,1})\sum\limits_{i=1}^{N(t)}e^{T_i^3 + T_i}\xi_1(T_i) + \left(\sum\limits_{i=1}^{N(t)}e^{T_i^3 + T_i}\xi_1(T_i)\right)^2\right\}\notag\\
    &\hskip2cm + e^{-2(t^2 + t)}(\cos(t) + 3)\left\{ 2(y_{0,2} - x_{0,2})\sum\limits_{i=1}^{N(t)}e^{T_i^2 + T_i}\xi_2(T_i) + \left(\sum\limits_{i=1}^{N(t)}e^{T_i^2 + T_i}\xi_2(T_i)\right)^2\right\}
\end{align}
where the continuous part of the Lyapunov function $V^c(t,\qvect(t))$ is defined as
\begin{align}
    V^c(t,\qvect(t)) &\triangleq e^{-2(t^3 + t)}(\sin(t) + 3)(y_{0,1} - x_{0,1})^2 + e^{-2(t^2 + t)}(\cos(t) + 3)(y_{0,2} - x_{0,2})^2
\end{align}
and $\xvect^c(t) \triangleq \Phi(t,0)\xvect_0$ denotes the continuous part of the perturbed trajectory $\xvect(t)$.
We can again compute the error bound on $V(t,\qvect(t))$ using the argument of~\thm{shot_contract_linear}. 
Conditioning on the number of jumps by time $t$ being $N(t) = k$, we obtain:
\begin{align}\label{eq:2D_ltv_diag_bound}
    \Ebb_k[\norm{\yvect(t) - \xvect(t)}^2] \leq \frac{1}{2}\Ebb_k[\norm{\yvect_0 - \xvect_0}^2]e^{-\beta_st} + \frac{\kappa_s(\beta_s,0,t)}{2}
\end{align}
where $\beta_s \triangleq 2\alpha = 4$ and $\kappa_s$ is as in~\eqn{kappa_shot_linear} with $\psi_k(0,t)$ defined as~\eqn{psi_ebb}. 
We use the simplification of $d\psi_k/dt$ using the maximum bound~\eqn{max_max_psi}, and substitute $k=10, \alpha_2 = 4, \beta = 1, \kappa = 2$.
The computation follows similarly to that of Section~\ref{subsec:1D_lqr} and is not repeated here for conservation of space.
The 1) evolution of $V$, $V^c$, the theoretical upper bound on $\Ebb_{10}[\norm{\yvect(t) - \xvect(t)}]$, and 2) evolution of the trajectories $\xvect$, $\xvect^c$, and $\yvect$ over time are visualized in~\fig{2D_diagonal_ltv_stable_bddP}.

\paragraph{Triangular $A(t)$ Matrix}
Now consider instead the case where $A(t)$ is an upper-triangular matrix, i.e., $a_{21}(t)\equiv 0$. Specifically, let
\begin{align}
    A(t) = \begin{bmatrix} -\cos(t) - 5 & 10\\ 0 & -2t - 1 \end{bmatrix}
\end{align}
which has state-transition matrix
\begin{align}
    \Phi(t,s) &\triangleq 
    \begin{bmatrix}e^{(\sin(t) - 5t) - (\sin(s) - 5s)} & 10(t-s)e^{(\sin(t) - 5t) - (\sin(s) - 5s)}\\ 0 & e^{-(t^2+t) + (s^2+s)}\end{bmatrix}
\end{align}

This allows us to write the solution trajectories~\eqn{shot_sde_linear_soln} for the perturbed ($\xvect(t)$) and nominal ($\yvect(t)$) systems explicitly as follows:
\begin{align}\label{eq:2D_tri_ltv_stable_bddP_solns}
    \xvect(t) &= \begin{bmatrix}
        e^{b_1(t)}(x_{0,1} + 10tx_{0,2})\\ e^{b_2(t)}x_{0,2}
    \end{bmatrix} + \sum\limits_{i=1}^{N(t)}\begin{bmatrix}
        e^{b_1(t) - b_1(T_i)}(\xi_1(T_i) + 10(t-T_i)\xi_2(T_i))\\ e^{b_2(t) - b_2(T_i)}\xi_2(T_i)
    \end{bmatrix}, \quad
    \yvect(t) = \begin{bmatrix}
        e^{b_1(t)}(y_{0,1} + 10ty_{0,2})\\ e^{b_2(t)}y_{0,2}
    \end{bmatrix}
\end{align}
where $b_1(t) \triangleq \sin(t) - 5t$ and $b_2(t) \triangleq -(t^2 + t)$.

\begin{figure}
    \begin{center}
    \includegraphics[width=0.6\columnwidth]{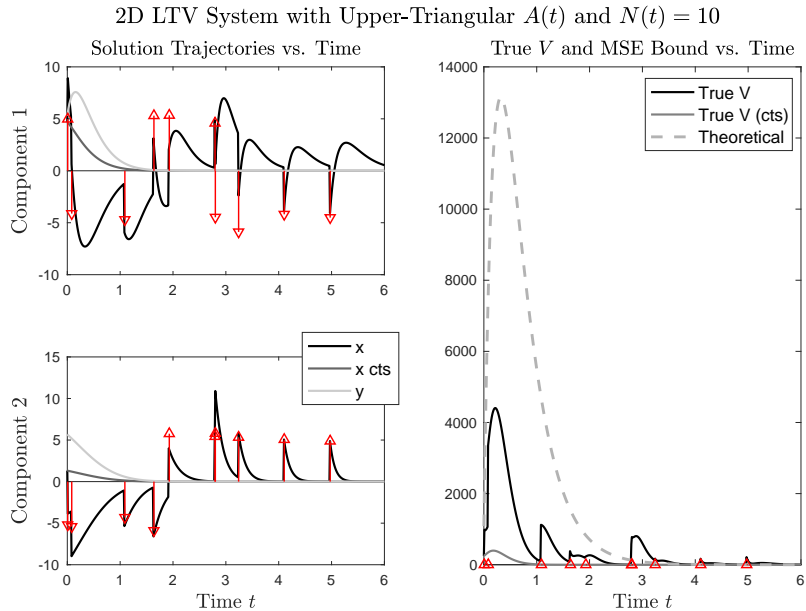}
    \caption{\fig{2D_diagonal_ltv_stable_bddP} for the exponentially stable 2D upper-triangular LTV system with bounded $P(t)$ metric. The theoretical error bound is computed with $d\psi_k/dt$ as in~\eqn{sum_exp_psi}.
    The number of jumps is again fixed at $N(t) = 10$.}
    \label{fig:2D_triangular_ltv_stable_bddP}
    \end{center}
\end{figure}

In contrast to the previous diagonal $A(t)$ case, we now consider a symmetric $P(t)$ with nonzero off-diagonal elements.
Namely:
\begin{align}
    P(t) \triangleq \begin{bmatrix} \sin(t)+3 & 1\\ 1 & \cos(t) + 3 \end{bmatrix}
\end{align}
which satisfies~\eqn{riccati_tv} with $\alpha=2$. This choice of $P(t)$ is also uniformly positive definite for each $t > 0$, and $P(t)$ is bounded as in~\assum{riccati_tv} with $\alpha_2 = 4.7071$ and $\alpha_1 = 1.2929$. We can now construct the Lyapunov-like function~\eqn{sde_lyap_linear} for this particular system. Compared to the previous diagonal matrix case, the inclusion of cross-terms make computation a little trickier.
\begin{align}
    V(t,\qvect(t)) &\triangleq (\yvect(t) - \xvect(t))^TP(t)(\yvect(t) - \xvect(t))\notag\\
    &= V^c(t,\qvect) - 2(\sin(t)+3)z_1(t)d_1(t) - 2(\cos(t)+3)z_2(t)d_2(t)\notag\\
    &\hskip1cm - 2\left[z_2(t)d_1(t) + z_1(t)d_2(t) - d_1(t)d_2(t) \right] + (\sin(t)+3)d_1^2(t) + (\cos(t)+3)d_2^2(t)
\end{align}
where
\begin{align}
    V^c(t,\qvect(t)) &\triangleq \left(\yvect(t) - \xvect^c(t)\right)^TP(t)\left(\yvect(t) - \xvect^c(t)\right) = (\sin(t) + 3)z_1^2(t) + 2z_1(t)z_2(t) + (\cos(t)+3)z_2^2(t)
\end{align}
with $\xvect^c(t) \triangleq \Phi(t,0)\xvect_0$ being the continuous part of the perturbed solution $\xvect(t)$, and
\begin{align*}
     z_1(t) \triangleq e^{b_1(t)}\left(y_{0,1} + 10ty_{0,2} - x_{0,1} - 10tx_{0,2}\right), &\quad z_2(t) \triangleq e^{b_2(t)}\left(y_{0,2} - x_{0,2}\right)\\
     d_1(t) \triangleq \sum\limits_{i=1}^{N(t)}e^{b_1(t) - b_1(T_i)}\left(\xi_1(T_i) + 10(t-T_i)\xi_2(T_i)\right), &\quad d_2(t) \triangleq \sum\limits_{i=1}^{N(t)} e^{b_2(t) - b_2(T_i)}\xi_2(T_i)
\end{align*}
Instead of dealing with this explicit version of the Lyapunov-like function, we can use the bound from~\thm{shot_contract_linear}. 
When the induced $2$-norm is used, state-transition matrix $\Phi(t,s)$ is bounded for a choice of $\kappa = 5$ and $\beta = 0.5$. 
Conditioning on the number of jumps by time $t$ being $N(t) = k$, we obtain:
\begin{align}\label{eq:2D_ltv_tri_bound}
    \Ebb_k[\norm{\yvect(t) - \xvect(t)}^2] \leq \frac{1}{1.2929}\Ebb_k[\norm{\yvect_0 - \xvect_0}^2]e^{-\beta_st} + \frac{\kappa_s(\beta_s,0,t)}{1.2929}
\end{align}
where $\beta_s \triangleq 2\alpha = 4$ and $\kappa_s$ is as in~\eqn{kappa_shot_linear} with $\psi_k(0,t)$ defined as~\eqn{psi_ebb} with $d\psi_k/dt$ simplified using the sum of exponentials bound~\eqn{sum_exp_psi}.
Substitute $k=10, \alpha_2 = 4.7071, \beta = 0.5, \kappa = 5$ to obtain the bound.
The 1) evolution of $V$, $V^c$, the theoretical upper bound on $\Ebb_{10}[\norm{\yvect(t) - \xvect(t)}]$, and 2) the evolution of the trajectories $\xvect$, $\xvect^c$, and $\yvect$ over time are visualized in~\fig{2D_triangular_ltv_stable_bddP}.


\section{Conclusion}\label{sec:conclusion}
In this paper, we designed incremental stability criteria for nonlinear stochastic systems perturbed by two types of non-Gaussian noise, both characterized by impulsive jumps. 
The Shot Contraction Theorem (\thm{shot_contract}) was designed for compound Poisson shot noise systems of the form~\eqn{shot_sde} while the L\'{e}vy Contraction Theorem (\thm{levy_contract}) was designed for finite-measure L\'{e}vy noise systems of the form~\eqn{combined_sde}. 
In~\thm{shot_contract_linear}, a specialization of the Shot Contraction Theorem was presented for linear time-varying nominal dynamics of the form~\eqn{shot_sde_linear}.
All three theorems show that, under the condition that a finite number of jumps arise from the noise process over a finite interval of time, solution trajectories corresponding to different initial conditions and different realizations of the noise process converge exponentially to within a bounded error ball of each other in the mean-squared sense under certain practical boundedness conditions on the parameters of the noise process and contraction metric.
We've shown that the convergence rate $\beta_s$ for~\eqn{shot_sde} is equal to that of the nominal system $\dot{\yvect} = f(t,\yvect)$ because the shot noise system behaves exactly as the deterministic system in between consecutive jumps.
\remk{tradeoff} discusses properties of the error bound $\kappa_s(\beta_s,s,t)$ defined in~\eqn{kappa_shot}, and makes the claim that 1) larger jump norm bounds $\eta$ correspond to larger error bounds, and 2) shorter interarrival times between jumps correspond to error bounds which grow larger over a shorter horizon of time.
Furthermore, the convergence rate $\beta_{\ell}$ from~\eqn{beta_levy} and error bound $\kappa_{\ell}(\beta_{\ell},s,t)$ from~\eqn{kappa_levy} of the L\'{e}vy noise system~\eqn{combined_sde} are shown to be nearly direct sums of the parameters for the white noise system~\eqn{white_params} and the shot noise system~\eqn{shot_params}, which is similar to the implications of the L\'{e}vy-Khintchine theorem.
The numerical simulations of~\sec{simulation} demonstrate our results.
We first showed empirical mean-squared error bounds for the specific 2D nonlinear system of Section~\ref{subsec:2D_nonlinear} over three different intensities $\lambda$ of the shot noise process.
More specifically, the 1D simple linear reference-tracking shot noise system in Section~\ref{subsec:1D_lqr} illustrates the tradeoff of~\remk{tradeoff} by considering varying intensities $\lambda$. 
We also demonstrate how to derive analytical expressions for the theoretical $\kappa_s(\beta_s,s,t)$ in Sections~\ref{subsec:1D_lqr} and~\ref{subsec:2D_ltv}.
In particular, the two 2D LTV systems of Section~\ref{subsec:2D_ltv} demonstrate the computation of the theoretical bounds for more complex systems than studied in Section~\ref{subsec:1D_lqr}; we show the process of choosing the metric $P(t)$ and all the appropriate parameters values to satisfy the assumptions of~\thm{shot_contract_linear}.

We emphasize that the benefits of our work are two-fold.
First, the phenomenon of impulsive jumps in noise processes is understudied for nonlinear stochastic systems in the controls community compared to Gaussian white noise despite being equally prevalent and important for many applications. 
Second, by addressing the prerequisite problem of stability characterization for shot and L\'{e}vy noise systems, we establish the foundations to enable model-based design of stochastic controllers and observers that are robust to shot and L\'{e}vy noise. 
By considering a class of noise models broader than the Gaussian assumption, we can expand the capabilities of model-based synthesis procedures. 
Thus, instead of using an entirely model-free approach to handle non-Gaussian noise perturbations, we can use model-free approaches to merely supplement the enhanced model-based synthesis baseline.
This allows for a design procedure which consumes less training time and data.
As mentioned in~\remk{finite_horz} and~\remk{white_vs_shot_bounds}, by deriving stability theorems which are independent of the initial conditions, we enable a measure of how far the perturbed trajectory will deviate from the nominal within any local horizon of time $[s,t]$ for any $s < t$; this allows for the potential development of controllers and observers which are online and adaptive.
The probabilistic guarantee of the theoretical bounds also enables a smarter design for controllers and observers by including a predictive component, by using the probability as a measure of expectation on the number of jumps that may arise in fixed, future horizons of time.
For future work, we leverage the results of~\sec{stoch_contract} to propose a specific controller synthesis procedure for nonlinear stochastic systems perturbed by shot and L\'{e}vy noise.

\section*{Acknowledgments}
The authors wish to thank the anonymous reviewers for their valuable suggestions which significantly improved the quality of this paper.\\
The authors would like to thank John C. Doyle for the insights he provided as motivation for this work.


\subsection*{Financial Disclosure}
This material is based upon work supported by the National Science Foundation Graduate Research Fellowship under Grant No. DGE‐1745301.

\subsection*{Conflict of Interest}
The authors declare no potential conflict of interests.

\bibliography{wileyNJD-AMA}

\end{document}